\documentclass[
  journal=pasa,
  manuscript=research-paper, 
  year=2023,
  volume=,
]{cup-journal}  

\makeatletter
\renewcommand*{\cup@author@fnsymbol}[1]{\number#1}
\makeatother

\usepackage{graphicx}
\usepackage{overpic}
\usepackage{xspace}

\def\ha{H$\alpha$}

\def\hii{H~{\sc ii}}

\usepackage{microtype,siunitx,booktabs,physics, multicol}
\makeatletter
\DeclareRobustCommand{\HI}{%
  \mbox{H\check@mathfonts\fontsize\sf@size\z@\selectfont I}%
}
\DeclareRobustCommand{\HII}{%
  \mbox{H\check@mathfonts\fontsize\sf@size\z@\selectfont II}%
}
\makeatother

\title{The MAGPI Survey: Evidence for Non-Universal Resolved Dust Attenuation Relations Beyond the Local Universe}

\author{A. Mailvaganam}
\affiliation{School of Mathematical and Physical Sciences, Macquarie University, NSW 2109, Australia}
\alsoaffiliation{Astrophysics and Space Technologies Research Centre, Macquarie University, Sydney, NSW 2109, Australia}
\email[A. Mailvaganam]{anilkumar.mailvaganam@hdr.mq.edu.au}

\author{T. Zafar}
\affiliation{School of Mathematical and Physical Sciences, Macquarie University, NSW 2109, Australia}
\alsoaffiliation{Astrophysics and Space Technologies Research Centre, Macquarie University, Sydney, NSW 2109, Australia}

\author{P. Corcho-Caballero}
\affiliation{Kapteyn Astronomical Institute, University of Groningen, PO Box 800, 9700 AV Groningen, The Netherlands}
\author{S. Salim}
\affiliation{Indiana University, 727 East 3rd St. Swain West 318, Bloomington,IN 47405-7105, USA}

\author{Y. Koyama}
\affiliation{National Astronomical Observatory of Japan, 2-21-1 Osawa, Mitaka, Tokyo 181-8588, Japan}

\author{K. E. Harborne}
\affiliation{Department of Physics, Centre for Extragalactic Astronomy, Durham University, South Road, Durham DH13LE, UK}
\alsoaffiliation{Department of Physics, Institute for Computational Cosmology, Durham University, South Road, Durham DH13LE, UK}
\alsoaffiliation{ARC Centre of Excellence for All Sky Astrophysics in 3 Dimensions (ASTRO-3D), Australia}

\author{C. D. P. Lagos}
\affiliation{International Centre for Radio Astronomy Research (ICRAR), The University of Western Australia, Crawley, WA 6009, Australia}
\alsoaffiliation{ARC Centre of Excellence for All Sky Astrophysics in 3 Dimensions (ASTRO-3D), Australia}

\author{J. T. Mendel}
\affiliation{Research School of Astronomy and Astrophysics, Australian National University, Canberra, ACT 2611, Australia}
\alsoaffiliation{ARC Centre of Excellence for All Sky Astrophysics in 3 Dimensions (ASTRO-3D), Australia}

\author{E. Wisnioski}
\affiliation{Research School of Astronomy and Astrophysics, Australian National University, Canberra, ACT 2611, Australia}
\alsoaffiliation{ARC Centre of Excellence for All Sky Astrophysics in 3 Dimensions (ASTRO-3D), Australia}

\author{I. U. Aalia}
\affiliation{School of Mathematical and Physical Sciences, Macquarie University, NSW 2109, Australia}
\alsoaffiliation{Astrophysics and Space Technologies Research Centre, Macquarie University, Sydney, NSW 2109, Australia}

\author{S. Ayyappan}
\affiliation{School of Mathematical and Physical Sciences, Macquarie University, NSW 2109, Australia}
\alsoaffiliation{Astrophysics and Space Technologies Research Centre, Macquarie University, Sydney, NSW 2109, Australia}

\author{S. Barsanti}
\affiliation{Sydney Institute for Astronomy (SIfA), School of Physics, The University of Sydney, NSW 2006, Australia}

\author{A. J. Battisti}
\affiliation{International Centre for Radio Astronomy Research (ICRAR), The University of Western Australia, Crawley, WA 6009, Australia}
\alsoaffiliation{Research School of Astronomy and Astrophysics, Australian National University, Canberra, ACT 2611, Australia}
\alsoaffiliation{ARC Centre of Excellence for All Sky Astrophysics in 3 Dimensions (ASTRO-3D), Australia}

\author{J. Bland-Hawthorn}
\affiliation{Sydney Institute for Astronomy (SIfA), School of Physics, The University of Sydney, NSW 2006, Australia}
\alsoaffiliation{ARC Centre of Excellence for All Sky Astrophysics in 3 Dimensions (ASTRO-3D), Australia}

\author{I. Breda}
\affiliation{Instituto de Astrof\'{i}sica e Ci\^{e}ncias do Espaço, Centro de Astrof\'isica da Universidade do Porto, Rua das Estrelas, 4150-762 Porto, Portugal \label{IA-CAUP}}

\author{S. Carlson}
\affiliation{School of Mathematical and Physical Sciences, Macquarie University, NSW 2109, Australia}
\alsoaffiliation{Astrophysics and Space Technologies Research Centre, Macquarie University, Sydney, NSW 2109, Australia}

\author{Q.-H. Chen}
\affiliation{Research School of Astronomy and Astrophysics, Australian National University, Canberra, ACT 2611, Australia}
\alsoaffiliation{ARC Centre of Excellence for All Sky Astrophysics in 3 Dimensions (ASTRO-3D), Australia}

\author{L. Cortese}
\affiliation{International Centre for Radio Astronomy Research (ICRAR), The University of Western Australia, Crawley, WA 6009, Australia}
\alsoaffiliation{ARC Centre of Excellence for All Sky Astrophysics in 3 Dimensions (ASTRO-3D), Australia}

\author{B. Courtney-Barrer}
\affiliation{School of Mathematics and Physics, University of Queensland, Brisbane, QLD 4072, Australia}
\alsoaffiliation{ARC Centre of Excellence for All Sky Astrophysics in 3 Dimensions (ASTRO-3D), Australia}

\author{S. M. Croom}
\affiliation{Sydney Institute for Astronomy (SIfA), School of Physics, The University of Sydney, NSW 2006, Australia}

\author{S. Ellis}
\affiliation{Astrophysics and Space Technologies Research Centre, Macquarie University, Sydney, NSW 2109, Australia}
\alsoaffiliation{Australian Astronomical Optics, Macquarie University, Sydney, NSW 2109, Australia}

\author{C. Foster}
\affiliation{School of Physics, University of New South Wales, Sydney, NSW 2052, Australia}

\author{E. Gjergo}
\affiliation{School of Astronomy and Space Science, Nanjing University, Nanjing 210093, China}
\alsoaffiliation{Key Laboratory of Modern Astronomy and Astrophysics (Nanjing University), Ministry of Education, Nanjing 210093, People's Republic of China}

\author{K. Grasha}
\affiliation{Research School of Astronomy and Astrophysics, Australian National University, Canberra, ACT 2611, Australia}
\alsoaffiliation{ARC Centre of Excellence for All Sky Astrophysics in 3 Dimensions (ASTRO-3D), Australia}

\author{S. Gurung-Lopez}
\affiliation{Observatori Astron`omic de la Universitat de Val`encia, Ed. Instituts d'Investigaci{'o}, Parc Cient{'i}fic. C/ Catedr{'a}tico Jos{'e} Beltr{'a}n, n2, 46980 Paterna, Valencia, Spain}
\alsoaffiliation{Departament d'Astronomia i Astrof{'i}sica, Universitat de Val`encia, 46100 Burjassot, Spain}

\author{S. Mobina Hosseini}
\affiliation{Department of Physics, Shahid Beheshti University, P.O. Box 19839-69411, Tehran, Iran}

\author{S. Jeon}
\affiliation{Department of Astronomy and Yonsei University Observatory, 50 Yonsei-ro, Seodaemun-gu, Seoul 03722, Republic of Korea}

\author{X. Lyu}
\affiliation{School of Mathematical and Physical Sciences, Macquarie University, NSW 2109, Australia}
\alsoaffiliation{Astrophysics and Space Technologies Research Centre, Macquarie University, Sydney, NSW 2109, Australia}

\author{T. Mukherjee}
\affiliation{School of Mathematical and Physical Sciences, Macquarie University, NSW 2109, Australia}
\alsoaffiliation{Astrophysics and Space Technologies Research Centre, Macquarie University, Sydney, NSW 2109, Australia}

\author{M. Mun}
\affiliation{Institut d'Astrophysique de Paris, CNRS, Sorbonne Universit\'e, 98bis Boulevard Arago, 75014, Paris, France}

\author{Hye-Jin Park}
\affiliation{Research School of Astronomy and Astrophysics, Australian National University, Canberra, ACT 2611, Australia}
\alsoaffiliation{ARC Centre of Excellence for All Sky Astrophysics in 3 Dimensions (ASTRO-3D), Australia}

\author{Y. Peng}
\affiliation{Department of Astronomy, School of Physics, Peking University, Beijing 100871, China}
\alsoaffiliation{Kavli Institute for Astronomy and Astrophysics, Peking University, Beijing 100871, China}

\author{L. A. Porta}
\affiliation{Australian Astronomical Optics, Macquarie University, Sydney, NSW 2109, Australia}

\author{J. Prathap}
\affiliation{School of Mathematical and Physical Sciences, Macquarie University, NSW 2109, Australia}
\alsoaffiliation{Astrophysics and Space Technologies Research Centre, Macquarie University, Sydney, NSW 2109, Australia}

\author{A. Raidani}
\affiliation{School of Mathematical and Physical Sciences, Macquarie University, NSW 2109, Australia}
\alsoaffiliation{Astrophysics and Space Technologies Research Centre, Macquarie University, Sydney, NSW 2109, Australia}

\author{R. S. Remus}
\affiliation{Universit\"ats-Sternwarte, Fakult\"at f\"ur Physik, Ludwig-Maximilians-Universit\"at M\"unchen, Scheinerstr. 1, 81679 M\"unchen, Germany}
\alsoaffiliation{Centre for Astrophysics and Supercomputing, Swinburne University, John Street, Hawthorn VIC 3122, Australia}

\author{B. S. Salmasi}
\affiliation{School of Mathematical and Physical Sciences, Macquarie University, NSW 2109, Australia}
\alsoaffiliation{Astrophysics and Space Technologies Research Centre, Macquarie University, Sydney, NSW 2109, Australia}

\author{G. Sharma}
\affiliation{University of Strasbourg, CNRS UMR 7550, Observatoire astronomique de Strasbourg, F-67000 Strasbourg, France}
\alsoaffiliation{SISSA–International School for Advanced Studies, Via Bonomea 265, I-34136 Trieste, Italy}
\alsoaffiliation{UWC–University of the Western Cape, Department of Physics and Astronomy, Cape Town 7535, South Africa}
\alsoaffiliation{IFPU–Institute for Fundamental Physics of the Universe, Via Beirut, 2, 34151 Trieste, Italy}
\alsoaffiliation{INFN-Sezione di Trieste, via Valerio 2, I-34127 Trieste, Italy}

\author{Sarah M. Sweet}
\affiliation{Research School of Astronomy and Astrophysics, Australian National University, Canberra, ACT 2611, Australia}
\alsoaffiliation{School of Mathematics and Physics, University of Queensland, Brisbane, QLD 4072, Australia}

\author{Dian P. Triani}
\affiliation{Center for Astrophysics | Harvard \& Smithsonian, 60 Garden St, Cambridge, MA 02138, USA}

\author{L. M. Valenzuela}
\affiliation{Universit\"ats-Sternwarte, Fakult\"at f\"ur Physik, Ludwig-Maximilians-Universit\"at M\"unchen, Scheinerstr. 1, 81679 M\"unchen, Germany}

\author{G. van de Ven}
\affiliation{Department of Astrophysics, University of Vienna, T"urkenschanzstra{\ss}e 17, 1180 Vienna, Austria}

\author{Bodo L. Ziegler}
\affiliation{Department of Astrophysics, University of Vienna, T"urkenschanzstra{\ss}e 17, 1180 Vienna, Austria}

\received {dd Mmm YYYY}
\revised  {dd Mmm YYYY}
\accepted {dd Mmm YYYY}
\published{2026}


\usepackage{tabularx}
\usepackage{hyperref} 
\hypersetup{colorlinks,citecolor=blue,linkcolor=blue,urlcolor=blue}

\begin{document}

\defcitealias{mailvaganam_2026j}{Paper I}

\begin{abstract}

We study the spatially resolved relation between dust attenuation ($A_V$) and star formation rate surface density ($\Sigma_{\mathrm{SFR}}$) in galaxies from the MAGPI survey ($0.25 < z < 0.42$). Using Balmer-decrement-based attenuation maps for 178 galaxies, we investigate whether the locally calibrated resolved $A_V$--$\Sigma_{\mathrm{SFR}}$ relation remains valid at intermediate redshift by comparing MAGPI with the local relation measured from MaNGA. We find a clear positive correlation between $A_V$ and $\Sigma_{\mathrm{SFR}}$ in MAGPI, with systematically higher attenuation than in MaNGA at fixed $\Sigma_{\mathrm{SFR}}$. After matching galaxies in stellar mass ($M_{*}$) and offset from the star-forming main sequence ($\Delta$SFMS), MAGPI galaxies remain more attenuated than MaNGA galaxies at fixed $\Sigma_{\mathrm{SFR}}$. The attenuation excess is strongest for galaxies below the SFMS ($\Delta A_V \sim 0.40$ mag), weaker for galaxies on the SFMS ($\Delta A_V \sim 0.28$ mag), and minimal for galaxies above the SFMS ($\Delta A_V \sim 0.07$ mag). The dependence of the offset on $\Delta$SFMS suggests that nebular attenuation on kpc scales is regulated not only by local star formation activity, but also by the global evolutionary state of the host galaxy. Together, these results indicate that the resolved $A_V$--$\Sigma_{\mathrm{SFR}}$ relation is not universal, and that locally calibrated attenuation relations may not fully describe galaxies at intermediate redshift. This highlights the need for attenuation calibrations that account for galaxy population and redshift when interpreting spatially resolved galaxy properties.

\end{abstract}


\section{Introduction}

Dust attenuation ($A_V$) remains one of the primary uncertainties in measuring star formation across cosmic time, directly affecting the interpretation of nebular emission and inferred galaxy properties \citep{salim20}. This is especially important because star formation rates (SFRs) are commonly derived from nebular emission lines (i.e. H$\alpha$), requiring attenuation corrections \citep{Kennicutt_1998_law,Kennicutt_evans_2012}. Unlike extinction along a single line of sight, the effective attenuation measured in galaxies depends on both the dust column density and the relative geometry between dust, stars, and ionised gas \citep{witt92,calzetti94,draine11,tomicic17}. Spatially resolved measurements of $A_V$ are therefore important for understanding how dust obscuration varies within galaxies and how it affects derived quantities such as the star formation rate surface density ($\Sigma_{\mathrm{SFR}}$) \citep{Kreckel_2013, Barrera_2020, greener20}.

The Balmer decrement (BD), commonly measured from the H$\alpha$/H$\beta$ line ratio, provides a widely used estimate of nebular attenuation in star-forming \hii\ regions, with only weak dependence on gas temperature under typical Case B recombination conditions \citep{baker38,osterbrock06,garn10,groves12,nelson16}. BD-based attenuation measurements are widely used to correct emission-line fluxes and recover dust-corrected SFRs. However, applying this method beyond the local Universe is observationally challenging because H$\beta$ is faint and can be affected by underlying stellar absorption \citep{garn10,Koyama_2015,Steidel_2014,Reddy_2015}. Even when reliable BD measurements are available, the inferred $A_V$ depends not only on the amount of dust, but also on how dust, stars, and ionised gas are distributed relative to one another. It is therefore unclear whether empirical attenuation relations calibrated in the local Universe remain applicable across different galaxy populations and redshifts. This question is important because nebular attenuation has been used as an indirect tracer of gas surface density when direct molecular gas observations are unavailable \citep[e.g.,][]{Yesuf_2019, Barrera_2020}.

BD-based attenuation is sensitive to the local dust surrounding star-forming regions, including both diffuse ISM and birth-cloud components \citep{Charlot_2000, calzetti01, Wild11, greener20}. Observational studies have shown that nebular attenuation is linked to both global and local galaxy properties, including stellar mass ($M_{*}$), star formation activity, metallicity, and stellar mass surface density \citep{garn10,zahid13,reddy15,Grootes_2013,Abdurro_28,Li_2021,Wild11,Battisti_2016,Battisti_2022}. These correlations suggest that $A_V$ is regulated by the interplay between dust content, star-forming activity, and the underlying structure of the host galaxy.

Large integral field spectroscopy (IFS) surveys have made it possible to move beyond integrated measurements and examine attenuation within galaxies on kpc scales \citep{Sanchez12,bundy15,Croom2021,medling18,Battisti_2026}. This is important because global measurements mix regions with different star formation histories, dust columns, and ionisation conditions. Spatially resolved spectroscopy allows us to measure the relation between $A_V$ and local quantities such as $\Sigma_{\mathrm{SFR}}$ on (sub)kpc scales, providing a way to test whether attenuation is primarily governed by local star formation or also depends on the broader evolutionary state of the host galaxy.

In \citet{mailvaganam_2026j}, hereafter Paper I, we used spatially resolved spectroscopy from the Mapping Nearby Galaxies at Apache Point Observatory (MaNGA) survey \citep{bundy15} to show that local $\Sigma_{\mathrm{SFR}}$ is the strongest predictor of $A_V$ in nearby star-forming galaxies. This provides a locally calibrated empirical relation between nebular attenuation and star formation activity. Whether this resolved relation remains valid beyond the local Universe is currently unknown. In this work, we use MaNGA as a local benchmark to test whether the resolved $A_V$--$\Sigma_{\mathrm{SFR}}$ relation remains valid at intermediate redshift ($z\sim0.3$). If attenuation is governed primarily by local gas and star-forming conditions traced by $\Sigma_{\mathrm{SFR}}$, then the resolved $A_V$--$\Sigma_{\mathrm{SFR}}$ relation should remain universal across cosmic time. Conversely, if attenuation also depends on evolving galaxy-scale properties such as gas fraction, ISM structure, star formation efficiency, or dust geometry, then the relation may vary with redshift and galaxy evolutionary state.

The star-forming main sequence (SFMS), the tight correlation between galaxy SFR and $M_*$ at a given epoch, provides a useful framework for describing a galaxy's global star-forming state \citep{Noeske07,speagel14}. A galaxy's offset from the SFMS ($\Delta$SFMS) traces whether it is forming stars below, on, or above the typical rate at its $M_*$. Since position relative to the SFMS is linked to differences in star formation efficiency and quenching state \citep{Saintonge_2012, Saintonge_2016, medling18, Janowiecki_2020}, $\Delta$SFMS provides a useful proxy for the evolutionary state of the host galaxy. We therefore use $\Delta$SFMS as a global parameter to test whether the resolved $A_V$--$\Sigma_{\rm SFR}$ relation depends only on local star formation activity or is additionally modulated by the host galaxy's evolutionary state.

This dependence on global galaxy state is particularly relevant for nebular attenuation. Galaxies at different positions relative to the SFMS can differ in their gas depletion times, star formation efficiencies, and dust--star geometry, meaning that $A_V$ may not be set by local star formation activity alone. This is consistent with \citet{Koyama_2018}, who found that galaxies below the SFMS tend to show enhanced "extra" attenuation toward nebular regions at fixed $M_{*}$, suggesting that the relative geometry between stars, dust, and ionised gas may change during the quenching process.

However, it remains unclear whether a locally calibrated resolved attenuation relation should remain valid at intermediate redshift. Galaxies at $z\sim0.3$ have, on average, higher gas fractions and different star formation efficiencies than local galaxies \citep{Daddi_2010, Tacconi_2018, Saintonge_2017}. Since dust is closely connected to the gas content of galaxies, changes in the gas reservoir and star formation efficiency are expected to alter the effective attenuation measured from nebular emission, potentially shifting the resolved $A_V$--$\Sigma_{\mathrm{SFR}}$ relation \citep{Santini14, Magdis_2012}. In addition, the spatial distribution of dust and star-forming regions may differ at earlier cosmic times, changing the relative geometry between dust, stars, and ionised gas.

To test this directly, we use observations from the Middle Ages Galaxy Properties with Integral Field Spectroscopy survey \citep[MAGPI;][]{Foster21}, and compare them against the locally calibrated MaNGA relation derived in \citetalias{mailvaganam_2026j}. By applying an equivalent resolved analysis framework to MAGPI galaxies at intermediate redshift, we test whether the local $A_V$--$\Sigma_{\mathrm{SFR}}$ relation remains universal or varies with galaxy evolutionary state. We further test whether observational effects such as spatial resolution and detection limits can account for the observed differences.

The paper is organised as follows: in \S2, we describe the MAGPI dataset and outline the MaNGA comparison sample used in this study; in \S3, we present our methodology for constructing $A_V$ maps and deriving spatially resolved star formation rate surface densities; in \S4, we present the resolved $A_V$--$\Sigma_{\mathrm{SFR}}$ relation, including its dependence on global galaxy state and tests of observational effects; in \S5, we discuss the physical interpretation of the observed trends and their implications for dust attenuation and star formation; and in \S6, we summarise our conclusions. Throughout the paper, we adopt a flat $\Lambda$CDM cosmology with $\Omega_\Lambda=0.7$, $\Omega_m=0.3$, and $H_0=70$ km s$^{-1}$ Mpc$^{-1}$.

\section{Data: The MAGPI survey}
\label{sec:magpi}
The MAGPI survey was carried out as a European Southern Observatory (ESO) VLT/MUSE Large Program (program ID: 1104.B-0536), targeting 56 fields within the Galaxy And Mass Assembly \citep[GAMA;][]{driver11} G12, G15, and G23 fields. The survey comprises 60 primary MAGPI galaxies with $M_\ast$ $>7\times10^{10}M_\odot$ within $0.25<z<0.35$. We refer the reader to \citet{Foster21} for the aims and details of the survey.

The MAGPI survey uses the MUSE wide-field mode ($1'\times1'$ with a spatial sampling of $0.2''$ per pixel), covering the wavelength range of 4700--9350\,\AA. Ground-layer adaptive optics (GLAO) are used at VLT UT4 to correct the atmospheric seeing. This results in a gap in simultaneous wavelength coverage from 5805--5965\,\AA\ due to the use of the GALACSI sodium laser notch filter \citep{hartke20}. The total integration time of each of the 56 fields is 4.4\,hours consisting of six observing blocks of $2\times1230$\,s exposures. The depth and large field of view of MAGPI data enable the detection of foreground sources within the local Universe and distant background sources covering the redshift range $2.9<z<6.7$ \citep{Mukherjee_2026}.

Segmentation maps for each MAGPI field using their white-light images are produced using the \textsc{profound} package \citep{Robotham18}, which is a source detection and image analysis package. Two-dimensional (seeing-convolved) emission-line flux maps were also generated using these segmentation maps from continuum-subtracted spectra with \textsc{gist} \citep{Bittner_2020}, a \verb|Python| wrapper for \textsc{pPXF} \citep{Cappellari_2004, Cappellari_2017} and \textsc{GandALF} \citep{Sarzi_2017}. The stellar continuum modelling accounts for underlying Balmer absorption features before the nebular emission-line fluxes are measured from the continuum-subtracted spectra. See \citet{Battisti_2026} for further details on the generation of MAGPI-derived emission-line maps.

\section{Methods}\label{methods}

\subsection{Attenuation and SFR surface density maps}\label{sec:av_sfr_maps}

We construct spatially resolved $A_V$ and $\Sigma_{\mathrm{SFR}}$ maps for MAGPI galaxies following the methodology developed in \citetalias{mailvaganam_2026j} for the MaNGA sample. We summarise the main steps below, noting survey-specific modifications for the MAGPI data where applicable:

\begin{itemize}

    \item The MAGPI emission-line maps are corrected for Milky Way foreground extinction using the \citet{Schlafly2011} dust maps and the \citet{cardelli89} extinction law. 

    \item We require H$\alpha$ and H$\beta$ to have signal-to-noise ratio (SNR) $>3$. This threshold is lower than the SNR $>5$ criterion adopted in \citetalias{mailvaganam_2026j}, reflecting the higher redshift of the MAGPI sample and the lower typical SNR of H$\beta$.

    \item We then derive nebular attenuation maps from the Balmer decrement, $\mathrm{BD}_{\rm obs}=F(\mathrm{H}\alpha)/F(\mathrm{H}\beta)$, assuming Case B recombination with an intrinsic ratio of $\mathrm{BD}_{\rm intrinsic}=2.86$ \citep{osterbrock1989} and adopting the \citet{calzetti01} attenuation law.

    \item Spaxels with $\mathrm{BD}_{\rm obs}<\mathrm{BD}_{\rm intrinsic}$ are excluded, as these produce negative attenuation values. Such values can arise from measurement uncertainties, residual continuum-subtraction effects, or variations in local ISM conditions.

    \item Spatially resolved SFR maps are derived from the attenuation-corrected H$\alpha$ luminosity using the \citet{Kennicutt_1998_law} calibration, converted from a Salpeter to a \citet{chabrier03} initial mass function by dividing by 1.53 \citep{gunawardhana13,davies16}, as in \citetalias{mailvaganam_2026j}.

    \item Star-forming spaxels are selected using the Baldwin, Phillips \& Terlevich \citep[BPT;][]{bpt81} diagnostic diagram. We additionally require [OIII] and [NII] to have SNR $>3$ for the BPT classification. Following \citetalias{mailvaganam_2026j}, we retain only spaxels below the \citet{Kauffmann_2003b} demarcation, selecting regions dominated by star formation and excluding composite, AGN-, or shock-dominated regions.

    \item We compute $\Sigma_{\mathrm{SFR}}$ by dividing the SFR in each spaxel by its projected physical area, calculated using the MAGPI angular pixel scale of $0.2''$ and the angular diameter distance at the galaxy redshift.

    \item After applying the BPT and SNR selections, removing spaxels that produce negative attenuation values, and restricting the sample to $0.25 < z < 0.45$, the final analysis sample contains 178 MAGPI galaxies and 8224 spaxels.

\end{itemize}

\subsection{Global galaxy properties}
\label{sec:global_properties}

Global $M_{*}$ and redshifts for MAGPI galaxies are taken from the MAGPI ProSpect catalogue, with $M_{*}$ computed using \textsc{ProSpect} \citep{Robotham2020} and spectroscopic redshifts derived using \textsc{MARZ} \citep{Hinton_2016}. Global SFRs are computed by summing the dust-corrected SFRs of all BPT-selected star-forming spaxels in each galaxy.

\begin{figure}[ht]
    \centering
    
    \includegraphics[width=\linewidth]{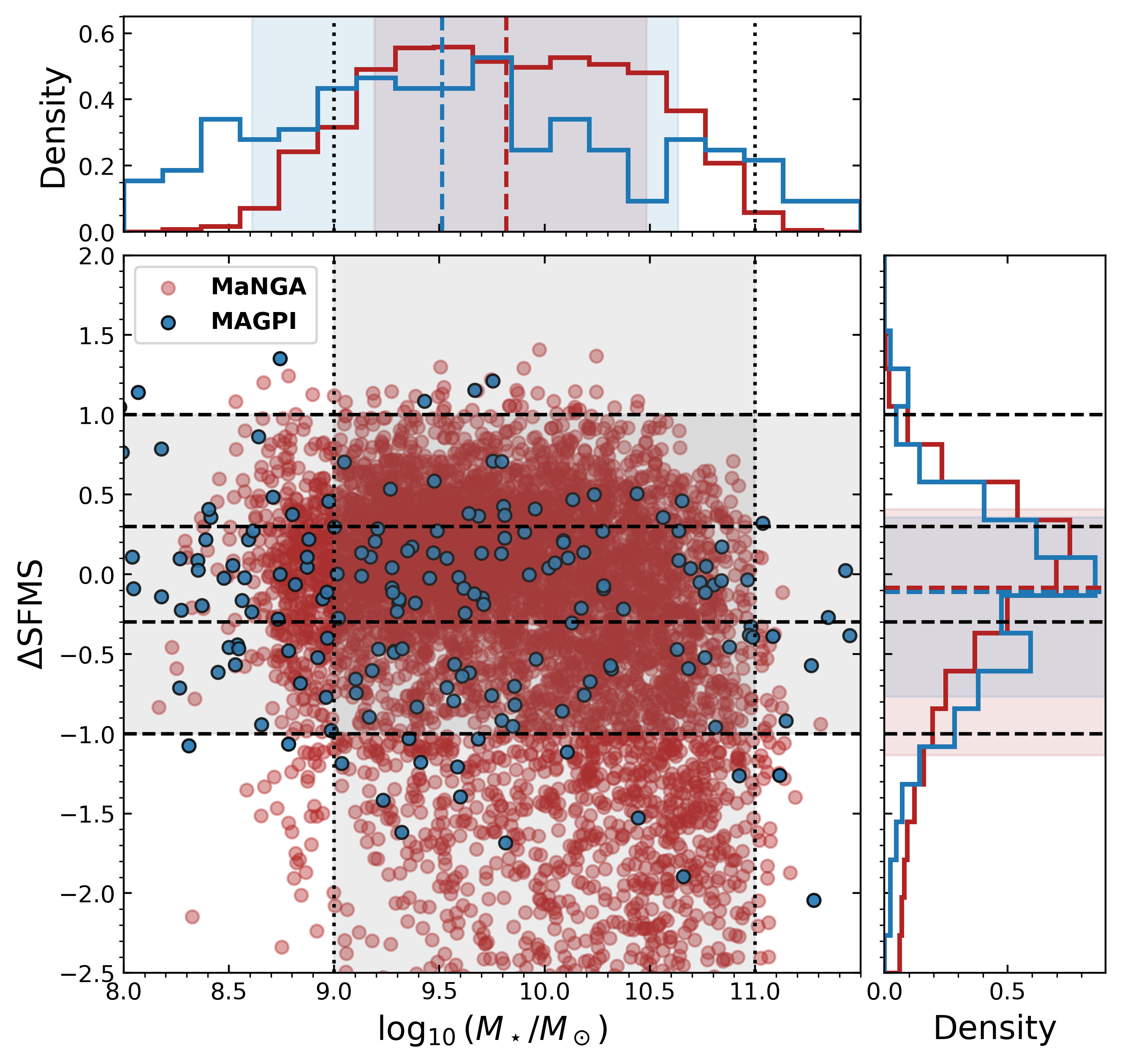}
    
    \caption{Distribution of MaNGA (red) and MAGPI (blue) galaxies in our sample in the stellar mass--$\Delta$SFMS plane. The top and right panels show the corresponding normalised distributions of stellar mass and $\Delta$SFMS, respectively, with dashed coloured lines indicating the medians and shaded regions showing the 16th--84th percentile ranges. Black dashed horizontal lines mark the adopted $\Delta$SFMS bin boundaries used to group galaxies below, on, and above the SFMS. Vertical dotted lines indicate the $M_{*}$ range used in the analysis. The grey shaded region highlights the parameter space adopted for the resolved comparison.}
    \label{fig:survey_distribution}
\end{figure}

\begin{figure}[ht]
    \centering
    
    \includegraphics[width=\linewidth]{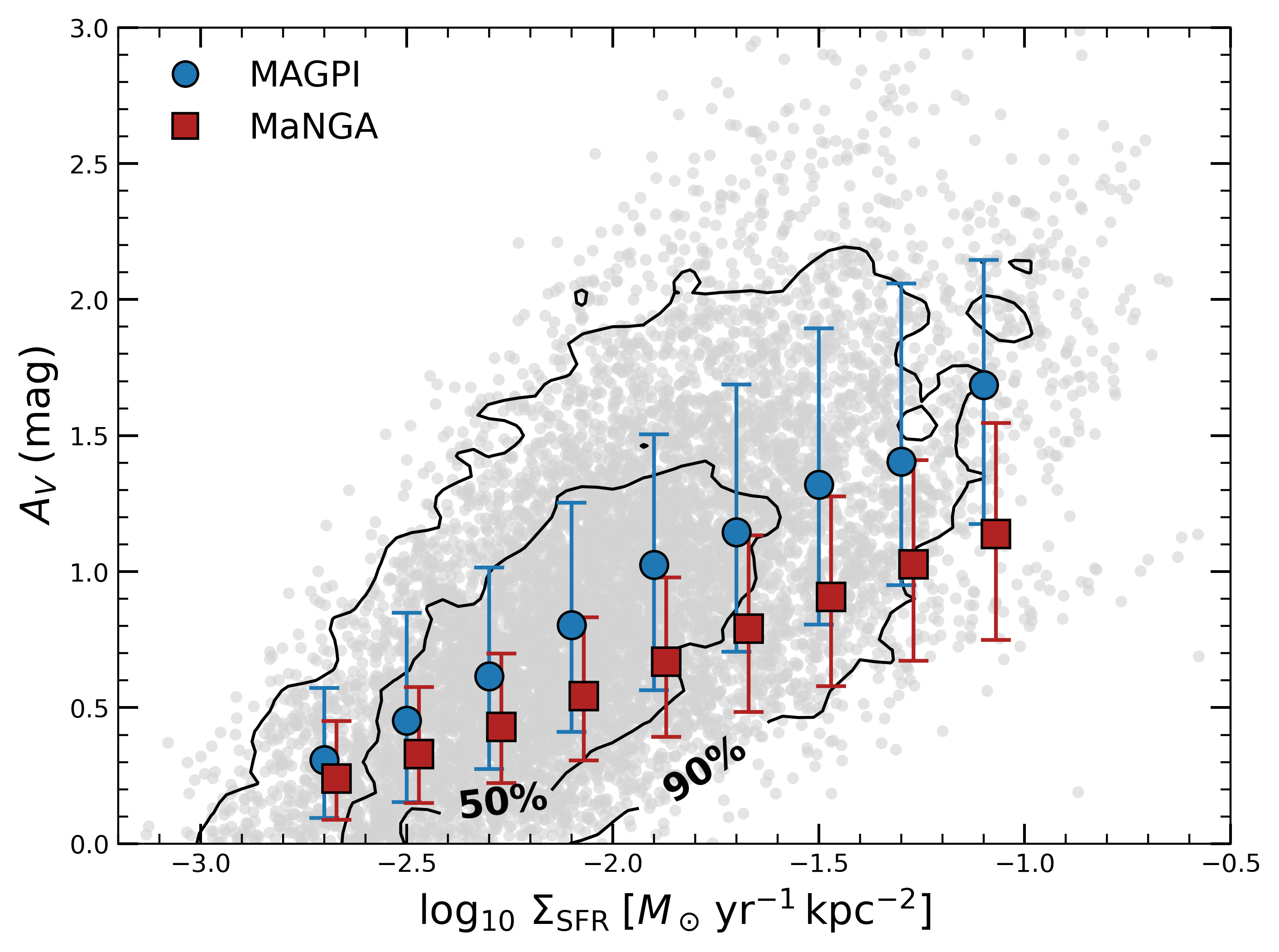}
    
    \caption{Relation between $A_V$ and $\log \Sigma_{\mathrm{SFR}}$ for the MAGPI sample. Grey points show individual MAGPI spaxels, with black contours indicating the 50\% and 90\% density levels of the MAGPI distribution. Blue circles show the median $A_V$ values in bins of $\log \Sigma_{\mathrm{SFR}}$, with error bars representing the 16th–84th percentile range. For comparison, red squares show the corresponding median relation measured from the MaNGA sample in the same $\log \Sigma_{\mathrm{SFR}}$ bins, with error bars indicating the 16th–84th percentile range.}

    \label{fig:Av_sfr_all}
\end{figure}

\begin{figure}[ht]
    \centering
    
    \includegraphics[width=\linewidth]{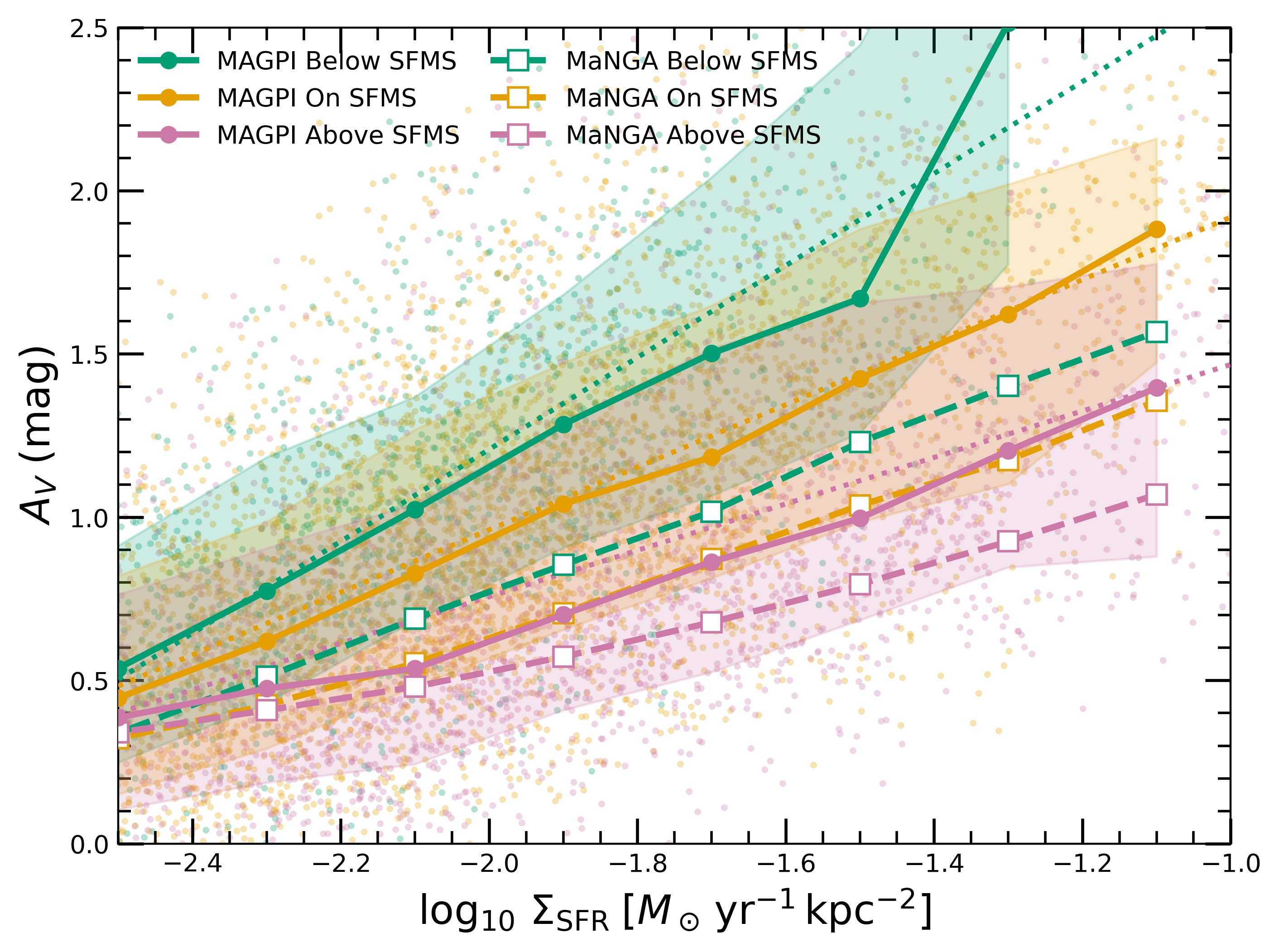}
    
    \caption{Relation between $A_V$ and $\log_{10}\Sigma_{\mathrm{SFR}}$ for the MAGPI and MaNGA samples, split by $\Delta$SFMS. The green, orange, and magenta colours show galaxies below, on, and above the SFMS, respectively. Individual points show MAGPI star-forming spaxels, while the solid curves show the median MAGPI $A_V$ in bins of $\log_{10}\Sigma_{\mathrm{SFR}}$. The shaded regions represent the 16th--84th percentile range of the MAGPI spaxel distribution within each bin. Dashed curves with open square markers show the corresponding median relations for the MaNGA sample, computed using the same binning and $\Delta$SFMS classification. Dotted lines show the best-fitting linear relations to the MAGPI spaxels in each $\Delta$SFMS bin, used to quantify the slope $a$ in Equation~\ref{eq:av_sfr_slope}}
    \label{fig:binned_SFMS}
\end{figure}

\begin{figure}[ht]
    \centering
    
    \includegraphics[width=\linewidth]{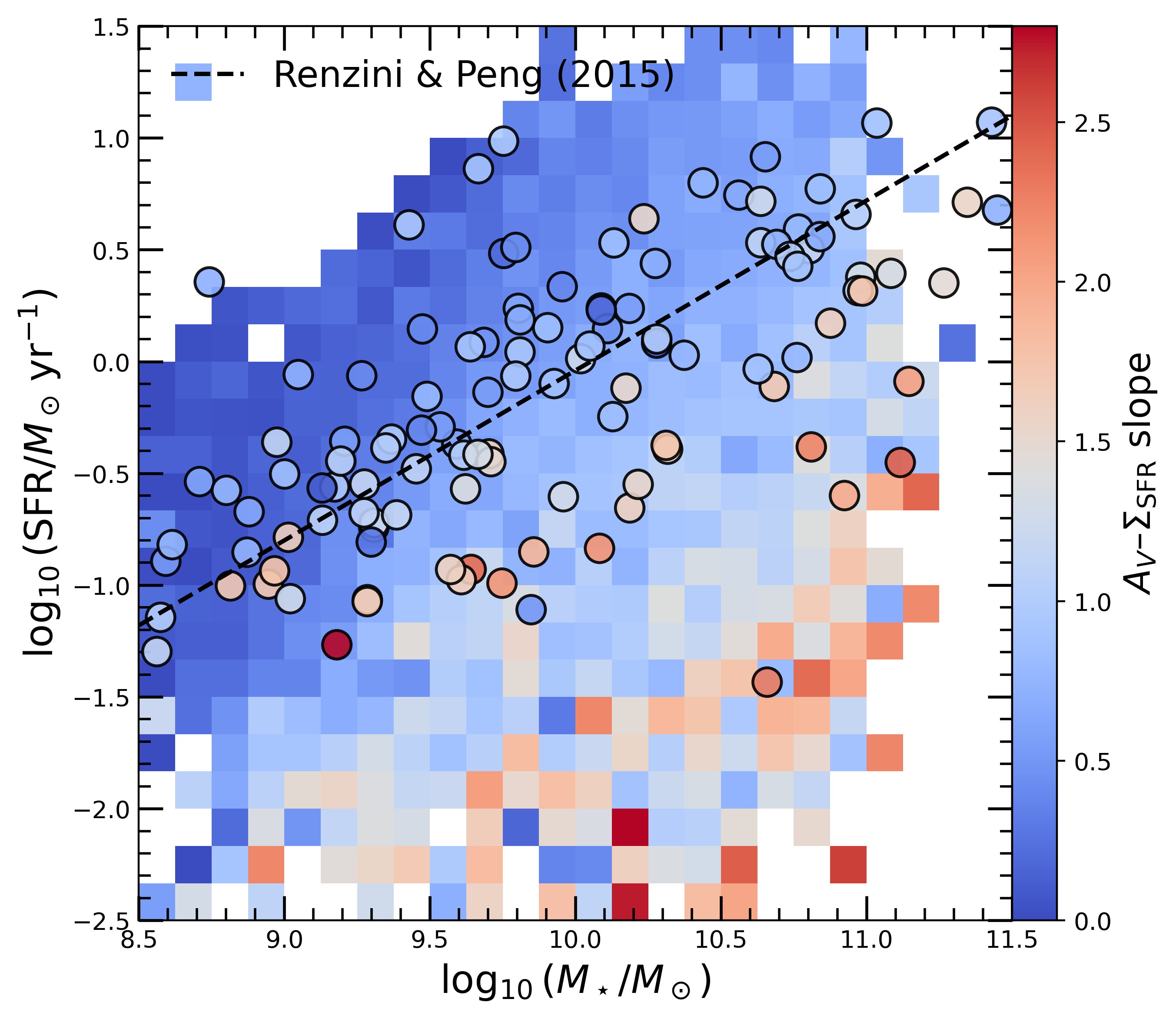}
    
    \caption{Distribution of MAGPI galaxies in the global SFR--$M_{*}$ plane, coloured by the fitted resolved $A_V$--$\log_{10}\Sigma_{\rm SFR}$ slope. Points show individual MAGPI galaxies, while the background map shows the median slope measured for MaNGA galaxies in bins of $\log_{10} M_{*}$ and $\log{\rm SFR}$. The dashed black line shows the star-forming main sequence from \citet{Renzini15}. }
    \label{fig:global_SFMS}
\end{figure}

\begin{figure*}
    \centering
    \includegraphics[width=\linewidth]{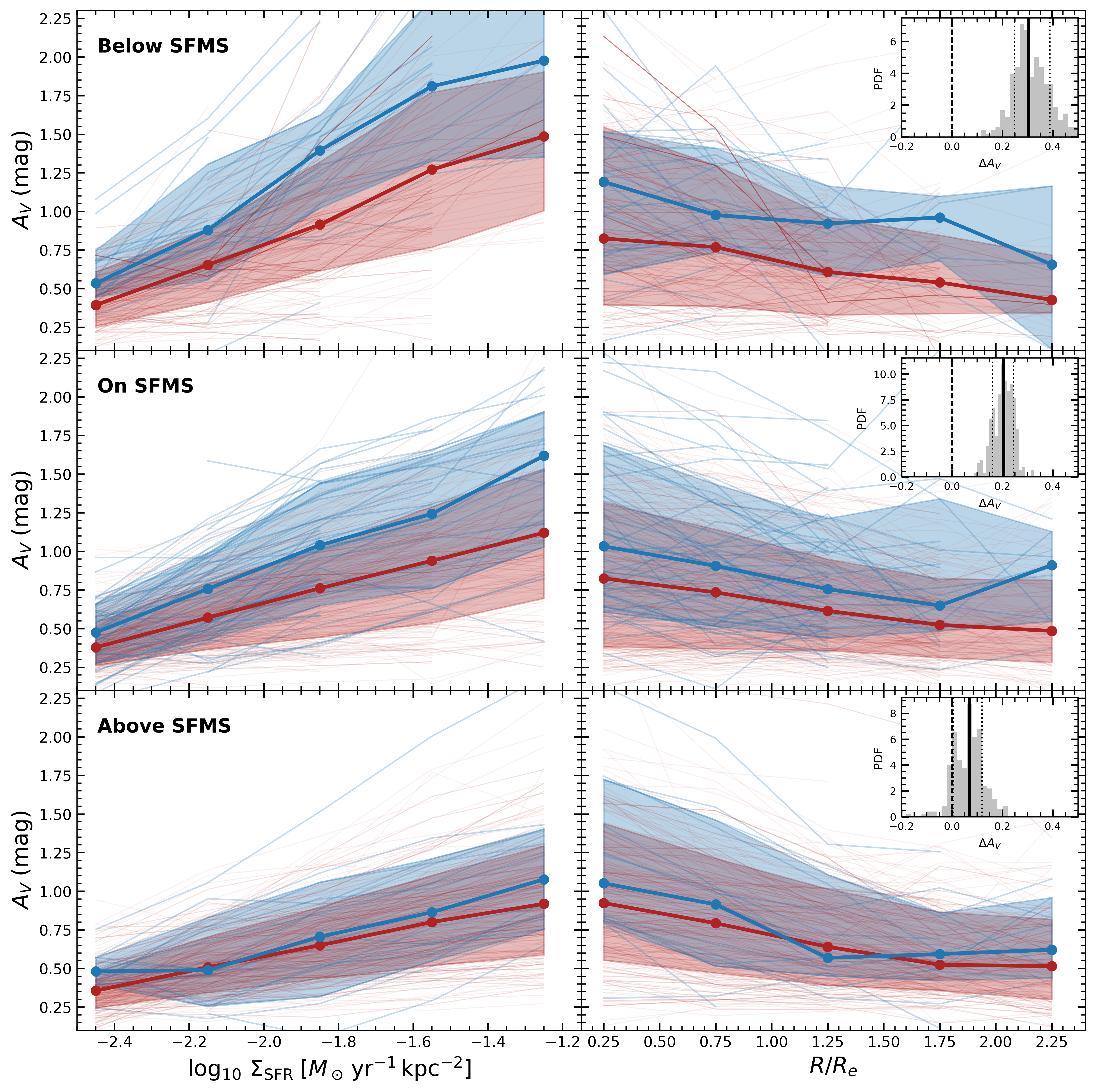}
    
    \caption{Comparison of the resolved $A_V$ profiles of MAGPI and matched MaNGA galaxies as a function of $\log_{10}\Sigma_{\mathrm{SFR}}$ (left column) and $R/R_e$ (right column), split by global $\Delta$SFMS. Rows show galaxies below, on, and above the SFMS, respectively. Blue curves represent MAGPI galaxies and red curves represent the matched MaNGA comparison sample. Thin curves show individual galaxy profiles, while thick curves with circular markers show the median relation. Shaded regions indicate the 16th--84th percentile range from the Monte Carlo matched samples. The inset histograms show the distribution of galaxy-level MAGPI--MaNGA attenuation offsets, $\Delta A_V = A_V^{\rm MAGPI} - A_V^{\rm MaNGA}$, for each $\Delta$SFMS bin. 
} 
    \label{fig:av_mcmc_snr_3}
\end{figure*}

\section{Results}

To provide a local benchmark, we compare the MAGPI analysis sample with nearby galaxies from the MaNGA survey ($z < 0.1$) analysed in \citetalias{mailvaganam_2026j}. While the MaNGA attenuation maps were originally constructed using an S/N$>5$ threshold, here we reconstruct the MaNGA $A_V$ and $\Sigma_{\mathrm{SFR}}$ maps using the same S/N$>3$ selection adopted for MAGPI. This ensures that both surveys are analysed under equivalent spaxel-selection criteria, minimising systematic differences arising from H$\beta$ sensitivity and detection limits.

\subsection{Global population differences}\label{population_differences}

We first compare the global stellar mass ($M_{*}$) and SFMS offset ($\Delta$SFMS) distributions of the MAGPI and MaNGA samples to assess population differences prior to the resolved analysis. Fig.~\ref{fig:survey_distribution} shows the distribution of galaxies in the $M_{*}$--$\Delta$SFMS plane for both surveys.

The vertical dotted lines indicate the stellar-mass range adopted for the resolved MAGPI--MaNGA comparison, $9 < \log_{10}(M_\star/M_\odot) < 11$, chosen to ensure overlap between the two surveys. Even within this common stellar-mass range, the stellar-mass distributions differ, with MaNGA exhibiting a broader distribution skewed toward higher masses compared to MAGPI.

The $\Delta$SFMS is computed relative to the star-forming main sequence (SFMS) relation from \citet{Renzini15}, and galaxies are grouped into three regimes: below the SFMS ($-1 < \Delta\mathrm{SFMS} < -0.3$), on the SFMS ($-0.3 < \Delta\mathrm{SFMS} < 0.3$), and above the SFMS ($0.3 < \Delta\mathrm{SFMS} < 1$).

MAGPI galaxies are shifted only slightly toward lower $\Delta$SFMS values, with a median of $-0.112$ compared to $-0.084$ for MaNGA. This median difference of $\sim0.03$ dex is smaller than the bootstrap uncertainty on the median difference ($\sim0.05$ dex), indicating that the global median $\Delta$SFMS values of the two samples are statistically consistent. However, within the adopted binning, MAGPI contains a larger fraction of galaxies below the SFMS (33\% versus 24\%) and a smaller fraction above the SFMS (15\% versus 28\%). Both surveys are dominated by galaxies on the SFMS, comprising 52\% of the MAGPI sample and 48\% of the MaNGA sample.

These differences suggest that the two surveys do not sample identical galaxy populations in global $(M_{*}, \Delta\mathrm{SFMS})$ space. As a result, any direct comparison of the resolved $A_{V}$--$\Sigma_{\rm SFR}$ relation between MAGPI and MaNGA may be influenced by population mixing. Possible environmental selection effects are considered in Section~\ref{environment_selection}.

\subsection{Overall resolved $A_V$--$\Sigma_{\rm SFR}$ relation}\label{full sample Av-ΣSFR}

We first compare the resolved relation between $A_V$ and $\Sigma_{\rm SFR}$ for the full MAGPI and MaNGA samples. Fig.~\ref{fig:Av_sfr_all} shows $A_V$ as a function of $\log \Sigma_{\rm SFR}$ for all star-forming spaxels, with median trends and corresponding percentile ranges overlaid for each survey. 

Both samples exhibit a clear positive correlation between $A_V$ and $\Sigma_{\rm SFR}$, consistent with previous studies linking nebular attenuation to local star formation activity \citep{Grootes_2013, Abdurro_28, Li_2021, Wild11}. In \citetalias{mailvaganam_2026j}, we showed that local $\Sigma_{\mathrm{SFR}}$ is strongly correlated with nebular attenuation in nearby star-forming galaxies, motivating the use of MaNGA as a local comparison sample.

However, a systematic offset is observed between the two surveys. At fixed $\Sigma_{\rm SFR}$, MAGPI spaxels exhibit higher $A_V$ compared to MaNGA across the full range of $\Sigma_{\rm SFR}$. The median offset between the MAGPI and 
MaNGA resolved relations is $\sim0.35$ mag across the overlapping 
$\Sigma_{\rm SFR}$ range, with bin-to-bin offsets spanning 
$\sim0.07$--$0.55$ mag. This offset is positive in all bins and is generally 
larger at higher $\Sigma_{\rm SFR}$.

The presence of this offset suggests that the $A_V$--$\Sigma_{\rm SFR}$ relation may depend on underlying differences in the galaxy populations sampled by each survey. In the following sections, we investigate whether this behaviour can be explained by variations in global galaxy properties.

\subsection{Resolved $A_V$--$\Sigma_{\rm SFR}$ relation as a function of global star-forming state}\label{SFMS_offset_dependence}

We further investigate the variation of the observed $A_V$--$\Sigma_{\rm SFR}$ relation in the MAGPI sample by grouping galaxies according to the $\Delta$SFMS bins defined in Section~\ref{population_differences}, as shown in Fig.~\ref{fig:binned_SFMS}. This approach allows us to separate galaxies by their star formation activity relative to the SFR expected for their $M_{*}$. Since $\Delta$SFMS measures a galaxy's offset from the mass-dependent SFMS, grouping galaxies by $\Delta$SFMS provides a way to compare systems by their relative global star-forming state \citep{Santini14}.

When grouped by the adopted $\Delta$SFMS bins, the MAGPI $A_V$--$\Sigma_{\rm SFR}$ relation shows a systematic variation with global star-forming state. At fixed $\Sigma_{\rm SFR}$, galaxies in the below-SFMS bin exhibit higher $A_V$ than those on or above the SFMS, while galaxies in the above-SFMS bin show the lowest attenuation. This indicates that the resolved $A_V$--$\Sigma_{\rm SFR}$ relation varies with position relative to the SFMS, rather than being determined by local $\Sigma_{\rm SFR}$ alone.

We quantify the $A_V$--$\Sigma_{\rm SFR}$ relation in each $\Delta$SFMS bin by fitting a linear model of the form
\begin{equation}
    A_V = a\,\log_{10}\left(\frac{\Sigma_{\rm SFR}}
    {M_\odot\,{\rm yr}^{-1}\,{\rm kpc}^{-2}}\right) + b ,
    \label{eq:av_sfr_slope}
\end{equation}
where $a$ is the fitted slope and $b$ is the intercept. These fits are shown as dotted lines in Fig.~\ref{fig:binned_SFMS}. We find that the slope decreases systematically from 
$a = 1.41 \pm 0.03$ for galaxies below the SFMS, to 
$a = 0.96 \pm 0.01$ for galaxies on the SFMS, and 
$a = 0.71 \pm 0.02$ for galaxies above the SFMS. The full best-fitting coefficients are listed in Table~\ref{tab:av_sfr_sfms_fits}. This indicates a progressive flattening of the resolved $A_V$--$\Sigma_{\rm SFR}$ relation with increasing global star formation activity.

Across the common $\Sigma_{\rm SFR}$ range sampled by the three $\Delta$SFMS bins, galaxies below the SFMS remain systematically more attenuated than those above the SFMS, with a typical separation of $\Delta A_V \sim 0.53$ mag. The scatter within the $\Delta$SFMS bins remains approximately consistent ($\sim 0.4$ mag), indicating that the observed differences reflect systematic variations in the relation rather than increased dispersion. This demonstrates that the resolved $A_V$--$\Sigma_{\rm SFR}$ relation depends strongly on global star-forming state.

The MaNGA median relations show a similar systematic trend with $\Delta$SFMS, with below-SFMS galaxies having higher $A_V$ at fixed $\Sigma_{\rm SFR}$ than galaxies on or above the SFMS. However, the separation between the $\Delta$SFMS bins is weaker than in MAGPI, and the MaNGA relations remain systematically lower in $A_V$, particularly below and on the SFMS. This suggests that the dependence of the resolved $A_V$--$\Sigma_{\rm SFR}$ relation on global star-forming state is also present locally, but is more pronounced in the MAGPI sample.

We further illustrate this trend in Fig.~\ref{fig:global_SFMS}, where the slope of the resolved $A_V$--$\Sigma_{\rm SFR}$ relation is mapped across the global $\log_{10} M_\star$--$\log_{10} \mathrm{SFR}$ plane. Each point represents an individual MAGPI galaxy, colour-coded by the best-fitting slope of its spaxel-level relation, while the background map shows the corresponding median slope per bin of $M_\ast$ and $\log_{10}\left({\rm SFR} / M_\odot\,yr^{-1}\right)$ for the MaNGA sample.

A clear gradient is observed relative to the star-forming main sequence, with galaxies below the SFMS exhibiting systematically steeper slopes, transitioning to flatter slopes toward and above the SFMS. This behaviour is consistent with the trends derived from the $\Delta$SFMS binning and demonstrates that the dependence of the $A_V$--$\Sigma_{\rm SFR}$ relation on global galaxy state is continuous across the population, rather than arising solely from discrete binning. In Fig.~\ref{fig:global_SFMS}, MAGPI galaxies also tend to exhibit steeper slopes than the corresponding MaNGA median slope values at similar locations in the $\log M_\star$--$\log \mathrm{SFR}$ plane, particularly in the low-SFR regime, with smaller differences toward higher SFR.

\begin{table}
\centering
\caption{
Best-fitting linear coefficients for the MAGPI resolved $A_V$--$\Sigma_{\rm SFR}$ relation \eqref{eq:av_sfr_slope} in each $\Delta$SFMS bin.
}
\label{tab:av_sfr_sfms_fits}
\begin{tabular}{lcc}
\hline
$\Delta$SFMS bin & $a$ & $b$ \\
\hline
$-1 < \Delta\mathrm{SFMS} < -0.3$ & $1.41 \pm 0.03$ & $4.02 \pm 0.07$ \\
$-0.3 < \Delta\mathrm{SFMS} < 0.3$    & $0.96 \pm 0.01$ & $2.88 \pm 0.03$ \\
$0.3 < \Delta\mathrm{SFMS} < 1$ & $0.71 \pm 0.02$ & $2.18 \pm 0.04$ \\
\hline
\end{tabular}
\end{table}

\subsection{Matched comparison between MAGPI and MaNGA}
\label{matched_comparison}

We investigate whether the resolved dust–SFR relation differs between the intermediate-redshift MAGPI sample and the local MaNGA sample by performing a matched analysis in $M_{*}$ and $\Delta$SFMS. This matching is motivated by the population differences identified in Section~\ref{population_differences}, and is used to reduce differences driven by the global galaxy-property distributions of the two surveys.

For each MAGPI galaxy, we identify candidate MaNGA counterparts within a tolerance of $\pm 0.1$ dex in both $\log M_{*}$ and $\Delta$SFMS. From this candidate set, a single MaNGA galaxy is randomly selected for each MAGPI galaxy in each realisation of the matching procedure.

This procedure is repeated over $N_{\rm MC} = 250$ Monte Carlo realisations to account for sampling variance in the MaNGA comparison sample. In each realisation, we compute the attenuation difference between matched galaxy pairs, $\Delta A_V = A_V^{\rm MAGPI} - A_V^{\rm MaNGA}$, using the binned median $A_V$ values measured as a function of $\Sigma_{\rm SFR}$ for each galaxy. For each matched pair, the attenuation difference is first computed in each $\Sigma_{\rm SFR}$ bin before being summarised as a single value by taking the median across $\Sigma_{\rm SFR}$ bins. The final attenuation offset is obtained by taking the median across all Monte Carlo realisations. We additionally derive resolved statistics to characterise how the offset varies with $\Sigma_{\rm SFR}$ and radius.

We then quantify the attenuation offset between the matched samples, as shown in Fig.~\ref{fig:av_mcmc_snr_3}, and find a systematic offset in $A_V$ between MAGPI and MaNGA galaxies after matching in $M_{*}$ and $\Delta$SFMS. Below the SFMS, the attenuation excess is strongest in MAGPI, with a median Monte Carlo offset of $\Delta A_V = 0.40_{-0.14}^{+0.13}$, where the uncertainties indicate the 16th--84th percentile range across matched realisations, and $\Delta A_V > 0$ in 100\% of Monte Carlo realisations. On the SFMS, the offset decreases to $\Delta A_V = 0.28_{-0.04}^{+0.05}$, but remains highly significant, with $\Delta A_V > 0$ in 100\% of realisations. Above the SFMS, the offset is much weaker, with $\Delta A_V = 0.07_{-0.06}^{+0.06}$ and $\Delta A_V > 0$ in 87\% of realisations, consistent with only a minimal difference in this regime. This offset varies systematically with $\Sigma_{\rm SFR}$, with MAGPI spaxels exhibiting higher $A_V$ than MaNGA at fixed $\Sigma_{\rm SFR}$ across the full dynamic range.

To test whether residual mismatches in global galaxy properties bias the observed $A_V$ offset, we examine the differences in $\log M_*$ and $\Delta$SFMS between matched MAGPI–MaNGA galaxy pairs across all Monte Carlo iterations. Both distributions are symmetric and centered around zero, indicating that the matching does not preferentially select galaxies on one side of the parameter range. We further test for any dependence of $\Delta A_V$ on these residual differences by comparing $\Delta A_V$ with $\Delta \log M_*$ and $\Delta$SFMS for all matched galaxy pairs. We find no correlation between $\Delta A_V$ and either quantity, indicating that small residual offsets in $M_*$ or SFMS position do not affect the measured offset.

To further investigate the spatial origin of the attenuation offset, we examine its radial dependence as a function of $R/R_{e}$, as shown in Fig.~\ref{fig:av_mcmc_snr_3}. This allows us to assess whether the offset is driven primarily by central regions or reflects a more radially extended difference between the two galaxy populations. $A_V$ is known to exhibit radial gradients in galaxies, with higher values typically found toward central regions and declining with increasing radius \citep[e.g.,][]{Nelson_2016, Gonzalez_Delgado_2015, greener20, Battisti_2026}. Below the SFMS, $\Delta A_V$ remains positive across the full radial range of $R/R_{e}$, with the excess particularly strong at both low and intermediate radii. On the SFMS, the offset also remains positive at all radii. In contrast, above the SFMS the radial behaviour is more mixed, with $\Delta A_V$ positive at low and high radii, but consistent with zero and mildly negative at intermediate radii around $R/R_e \sim 1.25$, indicating a weak and spatially varying difference in this regime.

\subsection{Effect of environmental selection}\label{environment_selection}

The MAGPI survey targets galaxies in fields centred on massive systems, and therefore may preferentially sample denser environments than the full MaNGA comparison sample. We test whether differences in environmental selection could contribute to the observed MAGPI--MaNGA attenuation offset.

To construct an environmentally selected MaNGA comparison sample, we use the X2 master catalogue from the GALEX--SDSS--WISE Legacy Catalog \citep[GSWLC-X2;][]{Salim_2016, Salim_2018} to identify massive SDSS galaxies with $M_\ast > 7\times10^{10}\,M_\odot$. For each MaNGA galaxy, we calculate the projected separation ($R_{\rm proj}$) from these massive GSWLC galaxies using their angular separation on the sky and the angular-diameter distance at the GSWLC redshift. We compute the line-of-sight velocity difference as
\begin{equation}
    |\Delta v| = c\,\frac{|z_{\rm MaNGA}-z_{\rm GSWLC}|}{1+z_{\rm GSWLC}},
\end{equation}
where $c$ is the speed of light. We retain MaNGA galaxies located within $R_{\rm proj}<1~\mathrm{Mpc}$ and $|\Delta v|<500~\mathrm{km\,s^{-1}}$ of at least one massive GSWLC galaxy, following similar projected-distance and velocity selections adopted in previous environment studies \citep[e.g.,][]{Kauffmann_2004}. This selection is therefore used as an environmental-proximity test for MaNGA galaxies near massive systems.

After applying the same selection cuts used in the main analysis, the environmentally selected MaNGA $A_V$--$\Sigma_{\rm SFR}$ relation is slightly elevated relative to the full MaNGA relation at fixed $\Sigma_{\rm SFR}$, with a median bin-wise increase of $\Delta A_V = 0.031$ mag and a bin-to-bin range of $0.018$--$0.053$ mag. This indicates that environmental selection may introduce a small upward shift in the local resolved attenuation relation.

However, this shift is not sufficient to remove the MAGPI--MaNGA offset. The median bin-wise MAGPI offset decreases from $\Delta A_V = 0.354$ mag relative to the full MaNGA sample to $\Delta A_V = 0.301$ mag relative to the environmentally selected MaNGA sample. Thus, differences in environmental selection change the inferred offset by only a small amount and are unlikely to be the dominant driver of the elevated attenuation observed in MAGPI galaxies at fixed $\Sigma_{\rm SFR}$.

\subsection{Effect of detection limits}\label{detection_limit}

Given the higher redshift of the MAGPI sample, a potential concern is that differences in detection limits could bias the comparison, as MAGPI may preferentially detect only the brightest H$\alpha$ regions. To test this, we apply a MAGPI-like H$\alpha$ luminosity threshold on the MaNGA spaxels, corresponding to the median detection limit derived for the MAGPI sample at $z \sim 0.30$.

The corresponding flux limit is estimated from the \ha\,error maps as 
$F_{\mathrm{H}\alpha,\mathrm{lim}} = 3 \times \sigma_{F,\mathrm{H}\alpha}$, 
where $\sigma_{F,\mathrm{H}\alpha}$ is the per-spaxel flux uncertainty. 
This is converted to a luminosity threshold via 
$L_{\mathrm{H}\alpha,\mathrm{lim}} = 4\pi D_L^2 F_{\mathrm{H}\alpha,\mathrm{lim}}$, 
evaluated at $z \sim 0.30$. We adopt a representative luminosity threshold of 
$L_{\mathrm{H}\alpha,\mathrm{lim}} \approx 5.6 \times 10^{37}\ \mathrm{erg\ s^{-1}}$, 
corresponding to the median value derived for the MAGPI sample.

This cut removes $\sim 39\%$ of the MaNGA spaxels, preferentially excluding low surface-brightness regions. As a result, the global median attenuation increases from $A_V = 0.59$ mag to $0.69$ mag, reflecting a change in the underlying population rather than the intrinsic relation. This preferentially removes regions more readily detected in MaNGA, making the test conservative.

At fixed $\Sigma_{\rm SFR}$, however, the $A_V$--$\Sigma_{\rm SFR}$ relation remains unchanged. As shown in Fig.~\ref{fig:detection_limit_test}, the median attenuation differs by at most $0.03$ mag across all $\Sigma_{\rm SFR}$ bins, with negligible variation over most of the dynamic range.

This variation is an order of magnitude smaller than the observed offset between MAGPI and MaNGA ($\sim 0.25$--$0.40$ mag), demonstrating that differences in detection limits cannot account for the systematic excess in $A_V$ observed in MAGPI. This indicates that the offset is not driven by observational sensitivity effects, but instead reflects an intrinsic difference between the galaxy populations.

We further test this by repeating the matched MAGPI--MaNGA comparison in Section~\ref{matched_comparison} after applying the same H$\alpha$ luminosity threshold to the MaNGA spaxels. The inferred attenuation offsets remain fully consistent with the original matched analysis, and increase slightly across all $\Delta$SFMS bins. At the spaxel level, the mean offset increases from $\Delta A_V = 0.25$ to $0.32$ below the SFMS, from $0.22$ to $0.26$ on the SFMS, and from $0.06$ to $0.07$ above the SFMS. At the global level, the offsets increase from $0.40$ to $0.47$ below the SFMS, from $0.28$ to $0.32$ on the SFMS, and from $0.07$ to $0.09$ above the SFMS. The radial $\Delta A_V$ profiles remain consistent within the uncertainties. Importantly, the attenuation offset is not reduced by imposing the detection limit, but instead becomes marginally stronger, demonstrating that the observed excess in $A_V$ is not driven by sensitivity effects.

\subsection{Effect of spatial resolution}\label{psf_effect}

Differences in physical spatial resolution can affect the recovery of spatially resolved emission-line properties in IFS data, especially when comparing surveys across redshift \citep{Mast_2014}. To assess whether the observed attenuation offset could be due to differences in spatial resolution between surveys, we performed a point spread function (PSF) matching test. Since our comparison is performed in physical units, we convolve the MaNGA emission-line maps with a Gaussian kernel to match the physical PSF of the MAGPI sample. The MAGPI physical PSF was estimated by converting the measured angular PSF for each galaxy into kpc using the angular diameter distance at the galaxy redshift. Based on the resulting MAGPI physical PSF distribution, which spans $\sim$1.9--2.9 kpc with a median of $\sim$2.3 kpc, we adopt a target resolution of $\sim$2.9 kpc, corresponding to the upper end of the MAGPI sample. The convolution is applied directly to the emission-line flux maps before recomputing $A_V$ and $\Sigma_{\rm SFR}$, ensuring a consistent physical comparison. Reflective boundary conditions are used to minimise artificial edge effects.

We find that PSF matching leads to only minimal changes in the measured attenuation relation, producing a mild flattening at low $\Sigma_{\rm SFR}$ consistent with beam smearing, but with a negligible impact on the overall behaviour.

At the spaxel level, the amplitude of the offset remains similar, with values of $\sim0.26$ below the SFMS, $\sim0.20$ on the SFMS, and $\sim0.03$ above the SFMS. At the global level, the attenuation offset also remains significant across all $\Delta$SFMS bins, with values of $\sim0.32$ below the SFMS, $\sim0.24$ on the SFMS, and $\sim0.05$ above the SFMS. The radial $\Delta A_V$ profiles remain consistent after PSF matching, with no significant change in their overall shape or amplitude.

These results demonstrate that while PSF effects introduce minor smoothing, they are not the dominant driver of the observed attenuation offset between MAGPI and MaNGA.

\section{Discussion}

Understanding whether locally derived attenuation relations remain valid across cosmic time is critical for interpreting spatially resolved star formation in galaxies. In this work, we use the MAGPI sample to test whether the MaNGA calibrated $A_V$--$\Sigma_{\rm SFR}$ relation, established in the local Universe, also applies to galaxies at intermediate redshift.

As shown in Section~\ref{full sample Av-ΣSFR}, the MAGPI and MaNGA samples do not follow the same resolved $A_V$--$\Sigma_{\rm SFR}$ relation. At fixed $\Sigma_{\rm SFR}$, MAGPI spaxels exhibit systematically higher $A_V$ than MaNGA spaxels, indicating that the locally calibrated relation does not fully describe galaxies at intermediate redshift.

This resolved offset is consistent with recent global measurements, where MAGPI galaxies were found to have higher BD values at fixed $M_{\ast}$ than galaxies in the local Universe \citep{Battisti_2026}, suggesting enhanced dust attenuation at intermediate redshift. This is broadly consistent with previous studies, finding increasing attenuation from the local Universe toward $z \sim 1$--2 \citep{Burgarella_2013, Andrews_2017, Zavala_2021} extending to $z \sim 5$ \citep{zafar19}.

$A_V$ is closely linked to the gas and dust content of galaxies, with dust generally tracing the underlying gas reservoir, while star formation is regulated by the Schmidt–Kennicutt relation between gas surface density and SFR \citep{Santini14, Kennicutt_1998_law}. At earlier cosmic times, galaxies are observed to have higher molecular gas fractions, which contribute to the elevated SFRs seen at higher redshift \citep{Daddi_2010, Tacconi_2018}. If the dust reservoir increases together with the gas reservoir, this can lead to higher dust columns and enhanced attenuation along a given line of sight \citep{Santini14, Magdis_2012}. This provides a possible explanation for the systematic offset observed in Fig.~\ref{fig:Av_sfr_all}, where MAGPI spaxels exhibit higher $A_V$ than MaNGA spaxels at fixed observed $\Sigma_{\rm SFR}$. However, it is important to note that BD-based $A_V$ is an effective attenuation measurement rather than a direct tracer of dust mass, since the connection between attenuation and dust content depends on the spatial distribution of dust and ionised gas \citep{Kreckel_2013}. The observed offset may therefore reflect not only higher dust columns, but also differences in the geometry or covering fraction of dust around star-forming regions.

As a further consistency check, we examine whether a similar trend is present within the MAGPI sample itself by dividing the galaxies into two redshift bins ($0.25 < z \leq 0.30$ and $0.30 < z \leq 0.42$). We then perform a matched comparison in $M_{\ast}$ and $\Delta$SFMS, following the same Monte Carlo approach described in Section~\ref{matched_comparison}. We find that higher-redshift MAGPI galaxies exhibit systematically higher $A_V$ at fixed $\Sigma_{\mathrm{SFR}}$, with the offset becoming more pronounced toward higher $\Sigma_{\mathrm{SFR}}$. This trend is consistent with the survey-level comparison between MAGPI and MaNGA, although the limited sample size and relatively small redshift range prevent a statistically robust measurement of evolution. The corresponding result is shown in Fig.~\ref{fig:av_evolution}.

Environmental selection provides another possible source of difference between the MAGPI and MaNGA samples, since MAGPI targets fields centred on massive systems. Environmental processes can suppress star formation in satellite galaxies \citep{Peng_2012, Wetzel2013}. In such cases, the current $\Sigma_{\rm SFR}$ may be reduced more rapidly than the dust and gas reservoirs contributing to nebular attenuation, producing slightly elevated $A_V$ at fixed $\Sigma_{\rm SFR}$. However, as shown in Section~\ref{environment_selection}, restricting the MaNGA comparison sample to galaxies in the vicinity of massive systems produces only a small upward shift in the local $A_V$--$\Sigma_{\rm SFR}$ relation and does not remove the attenuation offset. This suggests that environmental selection may introduce a small upward shift in the local relation, but is unlikely to be the dominant driver of the elevated $A_V$ observed in MAGPI at fixed $\Sigma_{\rm SFR}$.

The robustness tests presented in Sections~\ref{detection_limit} and~\ref{psf_effect} indicate that the MAGPI--MaNGA offset is unlikely to be driven primarily by sensitivity or spatial-resolution effects. In the detection-limit test, imposing a MAGPI-like H$\alpha$ luminosity threshold on MaNGA produces only minor changes to the resolved $A_V$--$\Sigma_{\rm SFR}$ relation and does not reduce the attenuation offset, including when the matched comparison is repeated after applying the threshold. Similarly, PSF matching the MaNGA emission-line maps to MAGPI-like physical resolution produces only minor changes to the relation. Together, these tests suggest that differences in sensitivity or spatial resolution are unlikely to account for the systematic $A_V$ excess observed in MAGPI, supporting the interpretation that the offset depends on global star-forming state.

As an additional robustness check, we repeated the analysis using a MAGPI-specific SFMS relation from \citet{Marcie_2024} to define $\Delta$SFMS for the MAGPI sample, while retaining the local SFMS relation from \citet{Renzini15} for MaNGA. The qualitative behaviour remains unchanged, with MAGPI galaxies still exhibiting elevated $A_V$ at fixed $\Sigma_{\mathrm{SFR}}$ relative to MaNGA, and the strongest offset occurring below the SFMS. This indicates that the observed attenuation offset is not driven by the specific choice of SFMS parameterisation, and is more likely to reflect differences in the underlying physical conditions of the two samples.

The dependence of the $A_V$--$\Sigma_{\rm SFR}$ relation on $\Delta$SFMS can be understood in terms of variations in star formation efficiency. Galaxies below the SFMS are generally characterised by longer gas depletion times and lower star formation efficiencies \citep{Tacconi_2018, Saintonge_2017}, which may reflect a stabilisation of the gas against collapse to form stars. This could arise from a reduced fraction of dense star-forming gas, with longer depletion times associated with suppressed gas fragmentation and a shift of the gas density distribution toward lower densities \citep{Saintonge_2012}. In this regime, the SFR can decline more rapidly than the underlying gas and dust reservoirs, leading to enhanced $A_V$ at fixed $\Sigma_{\rm SFR}$ and a partial decoupling between attenuation and local star formation. In contrast, galaxies above the SFMS may have larger gas reservoirs, shorter depletion times, or both \citep{Genzel_2015, Scoville_2016}, resulting in a closer coupling between $A_V$ and ongoing star formation. Galaxies on the SFMS likely lie between these regimes, where gas, dust, and star formation remain more tightly linked \citep{Lilly_2013}.

The stronger dependence of the $A_V$--$\Sigma_{\rm SFR}$ relation on $\Delta$SFMS in MAGPI compared to MaNGA suggests that the decoupling between attenuation and local star formation is more pronounced at intermediate redshift than in the local Universe. At earlier cosmic times, higher gas fractions may allow galaxies to retain larger gas and dust reservoirs relative to their instantaneous star formation activity, particularly as they move below the SFMS. In this regime, a decline in SFR may not be accompanied by an immediate reduction in the dust column traced by nebular emission, producing enhanced $A_V$ at fixed $\Sigma_{\rm SFR}$ and a steeper $A_V$--$\Sigma_{\rm SFR}$ relation, consistent with evidence that high-redshift quiescent galaxies can retain substantial gas and dust reservoirs while forming stars inefficiently \citep{Gobat_2018}. In contrast, local MaNGA galaxies are generally more gas-poor, which may reduce the magnitude of this effect and produce a flatter relation. The fact that the offset is strongest below the SFMS and weakens toward and above the SFMS suggests that the evolution of the resolved attenuation relation between the local and intermediate-redshift samples is amplified in regimes of reduced star formation efficiency.

The enhanced attenuation observed below the SFMS is also consistent with galaxies transitioning toward quiescence, where declining star formation is not immediately accompanied by a proportional reduction in the dust column toward the remaining nebular regions. In addition, the remaining star-forming regions may become more spatially distinct from the underlying stellar population. This is consistent with \citet{Koyama_2018}, who find that galaxies below the SFMS exhibit stronger extra attenuation toward nebular regions, quantified by the nebular-to-stellar colour excess ratio, $E(B-V)_{\rm gas}/E(B-V)_{\rm star}$, at fixed $M_{*}$. This suggests that nebular regions become increasingly distinct from the stellar continuum during quenching, as the remaining star formation becomes patchier or more centrally concentrated, or more deeply embedded within dusty birth clouds. Since BD-based attenuation primarily traces the local environments of H\,\textsc{ii} regions, an increased covering fraction or optical depth of birth clouds at intermediate redshift would naturally lead to elevated $A_V$ at fixed $\Sigma_{\rm SFR}$ \citep{Charlot_2000, Wild11}.

The radial $A_V$ profiles in Fig.~\ref{fig:av_mcmc_snr_3} show that the attenuation offset is not confined to galaxy centres, but persists across much of the radial extent of galaxies. This suggests that the MAGPI--MaNGA difference is not driven solely by central regions, but reflects a broader difference between the two samples. However, radial profiles do not directly constrain the physical distribution of dust. $A_V$ depends not only on the dust column, but also on the relative geometry between stars, dust, and ionised gas, as well as line-of-sight effects \citep{tomicic17, salim20, Koyama_2015}. Spatially resolved studies show that radial gradients in BD-based attenuation cannot be explained by diffuse ISM structure alone, but instead reflect a combination of local star formation and dust geometry \citep{greener20}.

This decoupling has direct consequences for derived physical quantities. In particular, applying a MaNGA-based attenuation relation to the MAGPI sample leads to systematic biases in the recovered $\Sigma_{\rm SFR}$, with the magnitude of the bias depending on global position relative to the SFMS. Galaxies below the SFMS show the largest deviations, while those above the SFMS are largely consistent with $\Delta \log \Sigma_{\rm SFR} \approx 0$. This suggests that the non-universality of the $A_V$--$\Sigma_{\rm SFR}$ relation can bias inferred star formation rates, particularly in galaxies below the SFMS, when locally calibrated attenuation relations are applied to intermediate-redshift samples.

These results have broader implications for spatially resolved studies of galaxies using IFU surveys. $A_V$ corrections are commonly applied to derive SFRs and to correct emission-line fluxes on a spaxel-by-spaxel basis \citep{Belfiore_2017, Chen_2023, medling18}. If locally calibrated attenuation relations are used in place of direct Balmer-decrement measurements, a systematic bias in the inferred $A_V$ can propagate into derived quantities such as $\Sigma_{\rm SFR}$, and can also affect emission-line diagnostics where dust corrections are required. This may influence the interpretation of resolved scaling relations and gradients. This is particularly relevant because $A_V$ is often used either as an input to dust-corrected SFRs or, in some studies, as an empirical tracer of gas content, including estimates of gas surface density and molecular gas mass when direct gas observations are unavailable \citep{Barrera_2020, Yesuf_2019}. Our results suggest that these $A_V$-based empirical calibrations should be applied carefully when extended beyond the local galaxy population, particularly for intermediate-redshift samples and galaxies below the SFMS.

\section{Conclusions}
\label{conclusion}

In this study, we have investigated the spatially resolved relation between $A_V$ and $\Sigma_{\mathrm{SFR}}$ for 178 galaxies from the MAGPI survey, and compared it to the corresponding relation measured in the local Universe using MaNGA \citepalias{mailvaganam_2026j}. Our aim was to test whether the locally derived $A_V$--$\Sigma_{\mathrm{SFR}}$ relation remains valid at intermediate redshift.

Our main findings are summarised as follows:

\begin{itemize}

    \item We find a clear positive correlation between $A_V$ and $\Sigma_{\mathrm{SFR}}$ in the MAGPI sample, consistent with the relation observed in MaNGA. However, at fixed $\Sigma_{\mathrm{SFR}}$, MAGPI galaxies show systematically higher attenuation than their local counterparts. The median resolved MAGPI relation is offset from MaNGA by $\sim0.35$ mag across the overlapping $\Sigma_{\mathrm{SFR}}$ range, with the offset generally larger at higher $\Sigma_{\mathrm{SFR}}$.

    \item The $A_V$--$\Sigma_{\mathrm{SFR}}$ relation in MAGPI depends strongly on global star-forming state. When splitting the sample by $\Delta$SFMS, we find that the fitted slope decreases from $a = 1.41 \pm 0.03$ below the SFMS, to $a = 0.96 \pm 0.01$ on the SFMS, and to $a = 0.71 \pm 0.02$ above the SFMS. At fixed $\Sigma_{\mathrm{SFR}}$, regions in galaxies below the SFMS exhibit significantly higher attenuation than those above the SFMS.

    \item After matching MAGPI and MaNGA galaxies in $M_{*}$ and $\Delta$SFMS, MAGPI galaxies remain more attenuated at fixed $\Sigma_{\rm SFR}$, showing that the offset is not driven solely by global population differences between the two surveys. The attenuation excess is strongest below the SFMS ($\Delta A_V \sim 0.40$ mag), decreases on the SFMS ($\Delta A_V \sim 0.28$ mag), and becomes minimal above the SFMS ($\Delta A_V \sim 0.07$ mag), indicating a strong dependence on global star-forming state.

    \item We demonstrate that observational and sample-selection effects are unlikely to drive this offset. Differences in detection limits and spatial resolution introduce only minor changes to the resolved relation. An environmental selection test shows that galaxies near massive systems have a slightly elevated local relation, but this shift is much smaller than the observed MAGPI excess. This suggests that the attenuation offset reflects differences in the $A_V$ properties of the two galaxy populations rather than being driven by observational or environmental biases.

\end{itemize}

The enhanced attenuation observed in MAGPI is consistent with differences in the coupling between gas, dust, and star formation relative to the local Universe. In particular, the stronger offset below the SFMS suggests that galaxies with reduced star formation efficiency may retain relatively large gas and dust reservoirs compared to their instantaneous star formation activity. While the limited redshift baseline prevents a direct measurement of cosmic evolution, our results suggest that locally calibrated attenuation relations may not fully capture the $A_V$ properties of intermediate-redshift galaxy populations.

Our findings have important implications for spatially resolved studies of galaxy evolution. Since $A_V$ corrections are routinely applied when deriving SFRs and other physical quantities, systematic differences in $A_V$ can propagate into inferred galaxy properties. This is particularly relevant for studies that rely on locally calibrated attenuation prescriptions, which may not fully capture the attenuation properties of galaxies below the SFMS at intermediate redshift.

Future work will extend this analysis to larger samples and wider redshift ranges to further test the robustness of the resolved $A_V$--$\Sigma_{\mathrm{SFR}}$ relation and its dependence on galaxy population.

\section{Acknowledgements}
CFoster is the recipient of an Australian Research Council Future Fellowship (project number FT210100168) funded by the Australian Government. 
\section{Data Availability}
The data underlying this article are available under the MAGPI survey. Additional data generated by the analyses in this work are available upon request to the corresponding author.


\bibliography{magpi_dust}

\appendix

\begin{table*}[p]
\caption{Properties of the MAGPI galaxies used in this work. The table lists the global quantities used to define $\Delta$SFMS, together with the fitted resolved $A_V$--$\log_{10}\Sigma_{\rm SFR}$ relation.}
\label{tab:magpi_av_sfr_slopes}
\centering
\resizebox{0.85\textwidth}{!}{%
\begin{tabular}{llllll}
\toprule
MAGPI ID & $z$ & $\log_{10}(M_\star/M_\odot)$ & $\log_{10}({\rm SFR}/M_\odot\,{\rm yr}^{-1})$ & $\Delta$SFMS & $A_V$--$\log_{10}\Sigma_{\rm SFR}$ slope \\
\midrule
2307145281 & 0.336 & 11.28 & -1.11 & -2.04 & 2.21 $\pm$ 0.31 \\
2305192319 & 0.385 & 10.66 & -1.43 & -1.89 & 1.62 $\pm$ 1.30 \\
1207136308 & 0.323 & 9.81 & -1.86 & -1.68 & 1.22 $\pm$ 0.02 \\
1522231213 & 0.293 & 9.32 & -2.18 & -1.62 & 7.03 $\pm$ 0.00 \\
1505245192 & 0.319 & 10.44 & -1.23 & -1.53 & 1.61 $\pm$ 0.23 \\
1523208089 & 0.327 & 9.23 & -2.04 & -1.42 & 2.35 $\pm$ 0.00 \\
1206305217 & 0.324 & 9.60 & -1.74 & -1.40 & 2.10 $\pm$ 0.28 \\
1524197197 & 0.330 & 10.92 & -0.60 & -1.26 & 2.08 $\pm$ 0.13 \\
1506325220 & 0.296 & 11.12 & -0.45 & -1.26 & 2.79 $\pm$ 0.97 \\
2311117362 & 0.381 & 9.59 & -1.56 & -1.21 & 3.08 $\pm$ 0.78 \\
1512293188 & 0.358 & 9.04 & -1.96 & -1.19 & 3.36 $\pm$ 0.66 \\
2301269332 & 0.299 & 9.41 & -1.67 & -1.18 & 2.14 $\pm$ 0.31 \\
2301257053 & 0.300 & 10.11 & -1.07 & -1.11 & 1.64 $\pm$ 0.26 \\
1503279150 & 0.289 & 8.31 & -2.40 & -1.07 & 2.21 $\pm$ 0.00 \\
2303073257 & 0.301 & 8.78 & -2.03 & -1.06 & 3.26 $\pm$ 0.65 \\
1528275314 & 0.397 & 9.68 & -1.31 & -1.03 & 1.21 $\pm$ 0.23 \\
2301167297 & 0.297 & 9.35 & -1.56 & -1.03 & 1.54 $\pm$ 0.79 \\
2311105068 & 0.370 & 8.99 & -1.79 & -0.98 & 2.50 $\pm$ 1.45 \\
1502225063 & 0.371 & 10.81 & -0.38 & -0.96 & 2.22 $\pm$ 0.23 \\
1206243216 & 0.324 & 9.85 & -1.11 & -0.95 & 0.51 $\pm$ 0.23 \\
1508232260 & 0.301 & 8.65 & -2.01 & -0.94 & 0.58 $\pm$ 0.93 \\
2304196198 & 0.286 & 11.15 & -0.09 & -0.92 & 2.04 $\pm$ 0.12 \\
1203287367 & 0.308 & 9.80 & -1.11 & -0.92 & 2.05 $\pm$ 0.32 \\
1528220154 & 0.321 & 9.17 & -1.57 & -0.89 & 0.68 $\pm$ 0.18 \\
2309160103 & 0.326 & 10.08 & -0.84 & -0.86 & 2.27 $\pm$ 0.24 \\
1501178229 & 0.306 & 9.39 & -1.33 & -0.83 & 1.43 $\pm$ 0.44 \\
1507232349 & 0.315 & 9.86 & -0.97 & -0.82 & 1.81 $\pm$ 0.22 \\
1509035197 & 0.373 & 9.57 & -1.16 & -0.80 & 1.08 $\pm$ 0.00 \\
1530072237 & 0.366 & 8.96 & -1.60 & -0.77 & 0.97 $\pm$ 0.08 \\
1530272149 & 0.308 & 9.75 & -0.99 & -0.76 & 2.12 $\pm$ 0.16 \\
1203191233 & 0.354 & 10.19 & -0.65 & -0.76 & 1.60 $\pm$ 0.45 \\
1509360309 & 0.284 & 9.10 & -1.46 & -0.74 & 1.45 $\pm$ 0.38 \\
2303323234 & 0.294 & 8.27 & -2.07 & -0.71 & 10.53 $\pm$ 0.00 \\
1509286279 & 0.420 & 9.53 & -1.10 & -0.71 & 1.18 $\pm$ 0.59 \\
1507190180 & 0.258 & 9.86 & -0.85 & -0.70 & 1.88 $\pm$ 0.25 \\
1512344182 & 0.321 & 8.84 & -1.60 & -0.68 & -0.19 $\pm$ 0.59 \\
\bottomrule
\end{tabular}
}
\end{table*}
\clearpage

\begin{table*}[p]
\caption{Continued.}
\label{tab:magpi_av_sfr_slopes_cont1}
\centering
\resizebox{0.85\textwidth}{!}{%
\begin{tabular}{llllll}
\toprule
MAGPI ID & $z$ & $\log_{10}(M_\star/M_\odot)$ & $\log_{10}({\rm SFR}/M_\odot\,{\rm yr}^{-1})$ & $\Delta$SFMS & $A_V$--$\log_{10}\Sigma_{\rm SFR}$ slope \\
\midrule
1505136036 & 0.316 & 10.22 & -0.55 & -0.67 & 1.53 $\pm$ 0.39 \\
1209119126 & 0.409 & 9.10 & -1.38 & -0.66 & 0.98 $\pm$ 0.38 \\
2302265123 & 0.294 & 9.61 & -0.98 & -0.64 & 1.75 $\pm$ 0.22 \\
1508152162 & 0.316 & 9.64 & -0.93 & -0.62 & 2.39 $\pm$ 0.34 \\
2305272316 & 0.372 & 8.45 & -1.83 & -0.61 & -0.46 $\pm$ 1.01 \\
1506111077 & 0.294 & 9.18 & -1.27 & -0.60 & 2.79 $\pm$ 0.46 \\
1531178256 & 0.348 & 10.32 & -0.39 & -0.59 & 1.62 $\pm$ 0.26 \\
1501224275 & 0.311 & 10.68 & -0.11 & -0.59 & 2.50 $\pm$ 0.64 \\
2304188222 & 0.283 & 10.31 & -0.38 & -0.57 & 1.80 $\pm$ 0.18 \\
1209197197 & 0.295 & 11.27 & 0.35 & -0.57 & 1.46 $\pm$ 0.08 \\
2308268103 & 0.346 & 8.53 & -1.72 & -0.57 & 2.75 $\pm$ 0.94 \\
2303206223 & 0.379 & 9.57 & -0.93 & -0.56 & 1.59 $\pm$ 0.30 \\
1205196165 & 0.292 & 9.96 & -0.60 & -0.53 & 1.65 $\pm$ 0.30 \\
2304158110 & 0.340 & 8.92 & -1.38 & -0.52 & 1.62 $\pm$ 0.71 \\
2309198195 & 0.326 & 10.76 & 0.02 & -0.52 & 0.77 $\pm$ 0.11 \\
2305332151 & 0.306 & 9.28 & -1.07 & -0.49 & 1.70 $\pm$ 0.25 \\
1531103073 & 0.347 & 9.28 & -1.07 & -0.48 & 1.06 $\pm$ 0.39 \\
1206291238 & 0.326 & 8.78 & -1.44 & -0.48 & 2.35 $\pm$ 0.33 \\
2301269327 & 0.298 & 10.63 & -0.03 & -0.47 & 0.87 $\pm$ 0.10 \\
2303268195 & 0.416 & 9.21 & -1.11 & -0.47 & 0.66 $\pm$ 0.58 \\
2311182091 & 0.332 & 9.32 & -1.02 & -0.47 & 1.26 $\pm$ 0.12 \\
2305094248 & 0.314 & 8.55 & -1.61 & -0.47 & 1.78 $\pm$ 0.63 \\
1203081107 & 0.306 & 8.50 & -1.64 & -0.46 & 1.53 $\pm$ 0.87 \\
2303197196 & 0.343 & 10.88 & 0.17 & -0.45 & 1.66 $\pm$ 0.09 \\
1203044142 & 0.366 & 8.54 & -1.59 & -0.44 & 0.41 $\pm$ 2.44 \\
1202300176 & 0.373 & 8.97 & -1.22 & -0.40 & 1.67 $\pm$ 0.69 \\
2304279199 & 0.285 & 10.99 & 0.32 & -0.39 & 2.01 $\pm$ 0.14 \\
1533176049 & 0.315 & 11.08 & 0.39 & -0.39 & 1.32 $\pm$ 0.10 \\
1207197197 & 0.321 & 11.45 & 0.68 & -0.38 & 0.66 $\pm$ 0.05 \\
1209131247 & 0.295 & 10.97 & 0.32 & -0.38 & 1.20 $\pm$ 0.09 \\
1204198199 & 0.316 & 10.98 & 0.38 & -0.33 & 1.31 $\pm$ 0.06 \\
2308295194 & 0.346 & 10.13 & -0.25 & -0.30 & 0.83 $\pm$ 0.22 \\
1502208255 & 0.295 & 8.73 & -1.28 & -0.28 & 1.83 $\pm$ 0.53 \\
1509152248 & 0.388 & 9.02 & -1.06 & -0.28 & 1.25 $\pm$ 0.47 \\
2308198197 & 0.347 & 11.35 & 0.71 & -0.27 & 1.42 $\pm$ 0.09 \\
1528310241 & 0.321 & 9.62 & -0.57 & -0.24 & 1.43 $\pm$ 0.19 \\
\bottomrule
\end{tabular}
}
\end{table*}
\clearpage

\begin{table*}[p]
\caption{Continued.}
\label{tab:magpi_av_sfr_slopes_cont2}
\centering
\resizebox{0.85\textwidth}{!}{%
\begin{tabular}{llllll}
\toprule
MAGPI ID & $z$ & $\log_{10}(M_\star/M_\odot)$ & $\log_{10}({\rm SFR}/M_\odot\,{\rm yr}^{-1})$ & $\Delta$SFMS & $A_V$--$\log_{10}\Sigma_{\rm SFR}$ slope \\
\midrule
1531314096 & 0.350 & 8.61 & -1.33 & -0.23 & 1.59 $\pm$ 0.59 \\
1523164082 & 0.389 & 9.30 & -0.81 & -0.23 & 0.28 $\pm$ 0.29 \\
2306085259 & 0.413 & 8.27 & -1.58 & -0.23 & 2.52 $\pm$ 0.16 \\
1506106169 & 0.294 & 10.37 & 0.03 & -0.22 & 0.78 $\pm$ 0.08 \\
1511160258 & 0.295 & 10.17 & -0.12 & -0.21 & 1.63 $\pm$ 0.15 \\
1207105205 & 0.339 & 8.37 & -1.47 & -0.20 & 0.81 $\pm$ 0.29 \\
2304045162 & 0.285 & 9.71 & -0.45 & -0.19 & 1.50 $\pm$ 0.17 \\
1523140218 & 0.357 & 9.39 & -0.69 & -0.18 & 1.15 $\pm$ 0.42 \\
1206262286 & 0.270 & 9.31 & -0.74 & -0.18 & 0.75 $\pm$ 0.16 \\
1525333203 & 0.322 & 8.56 & -1.30 & -0.16 & 1.14 $\pm$ 0.81 \\
1508336146 & 0.330 & 9.31 & -0.72 & -0.16 & 1.39 $\pm$ 0.22 \\
1208151070 & 0.371 & 8.94 & -1.00 & -0.15 & 1.70 $\pm$ 0.43 \\
1503316290 & 0.383 & 9.70 & -0.41 & -0.15 & 1.63 $\pm$ 0.24 \\
1209189246 & 0.295 & 8.18 & -1.56 & -0.14 & 1.33 $\pm$ 0.44 \\
2305342325 & 0.372 & 9.67 & -0.42 & -0.12 & 1.30 $\pm$ 0.34 \\
1511283231 & 0.294 & 10.76 & 0.43 & -0.11 & 0.87 $\pm$ 0.11 \\
1522094062 & 0.382 & 9.28 & -0.70 & -0.11 & 1.28 $\pm$ 0.24 \\
2308258220 & 0.348 & 8.97 & -0.94 & -0.11 & 1.74 $\pm$ 0.29 \\
1201129319 & 0.305 & 8.05 & -1.61 & -0.09 & 0.68 $\pm$ 0.25 \\
1204213219 & 0.254 & 10.28 & 0.08 & -0.09 & 0.09 $\pm$ 0.06 \\
2301288271 & 0.300 & 9.62 & -0.42 & -0.09 & 1.02 $\pm$ 0.17 \\
1512077133 & 0.320 & 9.27 & -0.68 & -0.08 & 1.10 $\pm$ 0.30 \\
1207181305 & 0.321 & 10.28 & 0.10 & -0.07 & 0.98 $\pm$ 0.07 \\
1503085308 & 0.382 & 10.80 & 0.50 & -0.07 & 1.14 $\pm$ 0.09 \\
1506083298 & 0.363 & 8.81 & -1.00 & -0.06 & 1.70 $\pm$ 0.28 \\
1506279269 & 0.297 & 10.74 & 0.47 & -0.05 & 1.09 $\pm$ 0.06 \\
2301064121 & 0.294 & 10.84 & 0.56 & -0.04 & 0.62 $\pm$ 0.12 \\
1530322331 & 0.367 & 10.96 & 0.66 & -0.03 & 1.14 $\pm$ 0.11 \\
1527156125 & 0.346 & 9.45 & -0.48 & -0.02 & 1.17 $\pm$ 0.29 \\
1203153287 & 0.307 & 8.48 & -1.22 & -0.02 & 1.14 $\pm$ 0.24 \\
1530089118 & 0.305 & 8.57 & -1.14 & -0.02 & 1.02 $\pm$ 0.35 \\
1203235348 & 0.360 & 9.59 & -0.37 & -0.02 & 0.61 $\pm$ 0.13 \\
1202221243 & 0.292 & 9.13 & -0.71 & -0.01 & 1.09 $\pm$ 0.15 \\
1530070238 & 0.366 & 9.93 & -0.10 & 0.00 & 0.86 $\pm$ 0.25 \\
2303148291 & 0.363 & 8.74 & -1.00 & 0.00 & 1.38 $\pm$ 0.16 \\
2308063334 & 0.387 & 9.01 & -0.79 & 0.00 & 1.63 $\pm$ 0.46 \\
\bottomrule
\end{tabular}
}
\end{table*}
\clearpage

\begin{table*}[p]
\caption{Continued.}
\label{tab:magpi_av_sfr_slopes_cont3}
\centering
\resizebox{0.85\textwidth}{!}{%
\begin{tabular}{llllll}
\toprule
MAGPI ID & $z$ & $\log_{10}(M_\star/M_\odot)$ & $\log_{10}({\rm SFR}/M_\odot\,{\rm yr}^{-1})$ & $\Delta$SFMS & $A_V$--$\log_{10}\Sigma_{\rm SFR}$ slope \\
\midrule
1528197197 & 0.322 & 11.43 & 1.07 & 0.02 & 0.88 $\pm$ 0.03 \\
2310293123 & 0.284 & 8.35 & -1.27 & 0.03 & 1.22 $\pm$ 0.44 \\
1503208231 & 0.289 & 10.69 & 0.52 & 0.04 & 0.76 $\pm$ 0.06 \\
1507150106 & 0.317 & 10.02 & 0.01 & 0.04 & 1.20 $\pm$ 0.15 \\
1512115127 & 0.322 & 9.27 & -0.55 & 0.04 & 1.15 $\pm$ 0.17 \\
1204331311 & 0.255 & 8.87 & -0.85 & 0.04 & 0.74 $\pm$ 0.16 \\
1203195161 & 0.279 & 10.77 & 0.59 & 0.05 & 0.80 $\pm$ 0.04 \\
1531086218 & 0.350 & 8.52 & -1.11 & 0.06 & 2.48 $\pm$ 0.80 \\
1534070145 & 0.308 & 10.05 & 0.07 & 0.07 & 0.83 $\pm$ 0.09 \\
1205197197 & 0.291 & 10.64 & 0.53 & 0.09 & 1.12 $\pm$ 0.06 \\
1507252203 & 0.394 & 8.35 & -1.20 & 0.09 & 1.34 $\pm$ 0.23 \\
1531338083 & 0.344 & 8.27 & -1.26 & 0.10 & 3.10 $\pm$ 0.81 \\
1206322202 & 0.326 & 9.54 & -0.29 & 0.10 & 0.56 $\pm$ 0.11 \\
1509291230 & 0.299 & 10.11 & 0.15 & 0.10 & 0.51 $\pm$ 0.09 \\
1201302222 & 0.299 & 9.17 & -0.56 & 0.11 & 0.97 $\pm$ 0.25 \\
2311173217 & 0.333 & 8.87 & -0.79 & 0.11 & 0.85 $\pm$ 0.10 \\
1501329201 & 0.272 & 8.04 & -1.42 & 0.11 & 0.54 $\pm$ 0.70 \\
2308076271 & 0.388 & 9.79 & -0.07 & 0.13 & 0.93 $\pm$ 0.13 \\
1502079084 & 0.341 & 9.70 & -0.14 & 0.13 & 0.55 $\pm$ 0.12 \\
2308129329 & 0.388 & 9.47 & -0.31 & 0.13 & 0.39 $\pm$ 0.15 \\
1534261263 & 0.369 & 9.13 & -0.57 & 0.14 & 0.10 $\pm$ 0.31 \\
1507211057 & 0.315 & 10.18 & 0.24 & 0.14 & 0.56 $\pm$ 0.05 \\
1534082214 & 0.315 & 9.35 & -0.39 & 0.15 & 0.99 $\pm$ 0.11 \\
1533197198 & 0.315 & 10.84 & 0.77 & 0.17 & 0.83 $\pm$ 0.07 \\
1507103239 & 0.315 & 9.37 & -0.35 & 0.17 & 0.95 $\pm$ 0.18 \\
2310167176 & 0.283 & 10.09 & 0.23 & 0.20 & 0.41 $\pm$ 0.22 \\
1531268070 & 0.344 & 9.19 & -0.45 & 0.21 & 1.05 $\pm$ 0.12 \\
1502071104 & 0.371 & 10.09 & 0.24 & 0.21 & 0.55 $\pm$ 0.18 \\
1201123284 & 0.349 & 8.59 & -0.89 & 0.22 & 0.53 $\pm$ 0.20 \\
2304119316 & 0.288 & 8.39 & -1.05 & 0.22 & 1.38 $\pm$ 0.33 \\
1534282147 & 0.318 & 8.88 & -0.67 & 0.22 & 0.69 $\pm$ 0.20 \\
1533161121 & 0.408 & 9.81 & 0.04 & 0.23 & 0.93 $\pm$ 0.10 \\
2308248215 & 0.348 & 9.90 & 0.15 & 0.26 & 0.81 $\pm$ 0.09 \\
1208109039 & 0.372 & 10.27 & 0.44 & 0.27 & 0.69 $\pm$ 0.09 \\
1202077074 & 0.398 & 9.49 & -0.16 & 0.27 & 0.79 $\pm$ 0.16 \\
2310199196 & 0.283 & 10.64 & 0.72 & 0.27 & 1.28 $\pm$ 0.05 \\
\bottomrule
\end{tabular}
}
\end{table*}
\clearpage

\begin{table*}[p]
\caption{Continued.}
\label{tab:magpi_av_sfr_slopes_cont4}
\centering
\resizebox{0.85\textwidth}{!}{%
\begin{tabular}{llllll}
\toprule
MAGPI ID & $z$ & $\log_{10}(M_\star/M_\odot)$ & $\log_{10}({\rm SFR}/M_\odot\,{\rm yr}^{-1})$ & $\Delta$SFMS & $A_V$--$\log_{10}\Sigma_{\rm SFR}$ slope \\
\midrule
1209328270 & 0.259 & 8.61 & -0.82 & 0.27 & 0.75 $\pm$ 0.15 \\
1529336091 & 0.293 & 9.21 & -0.36 & 0.29 & 0.57 $\pm$ 0.11 \\
1203247089 & 0.306 & 9.00 & -0.50 & 0.30 & 0.86 $\pm$ 0.18 \\
1202197197 & 0.291 & 11.04 & 1.07 & 0.32 & 0.86 $\pm$ 0.02 \\
1502258107 & 0.296 & 8.41 & -0.89 & 0.36 & 1.12 $\pm$ 0.26 \\
1534194077 & 0.313 & 10.56 & 0.74 & 0.36 & 0.69 $\pm$ 0.07 \\
1527173209 & 0.288 & 9.69 & 0.09 & 0.36 & 0.50 $\pm$ 0.07 \\
2311102063 & 0.370 & 9.81 & 0.19 & 0.37 & 0.67 $\pm$ 0.11 \\
1530093302 & 0.366 & 8.80 & -0.58 & 0.38 & 0.78 $\pm$ 0.30 \\
1529345176 & 0.288 & 9.64 & 0.07 & 0.38 & 0.90 $\pm$ 0.09 \\
2310137110 & 0.283 & 8.40 & -0.85 & 0.41 & 0.49 $\pm$ 0.09 \\
1523135170 & 0.320 & 9.95 & 0.33 & 0.41 & 0.39 $\pm$ 0.08 \\
1508217276 & 0.301 & 9.80 & 0.24 & 0.43 & 0.61 $\pm$ 0.07 \\
1524134045 & 0.324 & 8.97 & -0.36 & 0.46 & 1.13 $\pm$ 0.15 \\
2304104201 & 0.288 & 10.65 & 0.92 & 0.46 & 0.54 $\pm$ 0.03 \\
2301109255 & 0.294 & 10.13 & 0.53 & 0.47 & 0.86 $\pm$ 0.05 \\
1524292106 & 0.329 & 8.71 & -0.54 & 0.48 & 0.60 $\pm$ 0.18 \\
1508197198 & 0.316 & 10.24 & 0.64 & 0.50 & 1.52 $\pm$ 0.06 \\
1203076068 & 0.306 & 10.44 & 0.80 & 0.51 & 0.72 $\pm$ 0.06 \\
2308254223 & 0.348 & 7.86 & -1.14 & 0.53 & 2.04 $\pm$ 0.71 \\
2308186140 & 0.348 & 9.27 & -0.06 & 0.53 & 0.42 $\pm$ 0.12 \\
1506082320 & 0.294 & 9.47 & 0.15 & 0.59 & 0.40 $\pm$ 0.10 \\
2304202299 & 0.337 & 9.05 & -0.06 & 0.71 & 0.61 $\pm$ 0.11 \\
1507084083 & 0.258 & 9.80 & 0.51 & 0.71 & 0.37 $\pm$ 0.05 \\
1206151090 & 0.325 & 9.75 & 0.48 & 0.71 & 0.16 $\pm$ 0.06 \\
1209257187 & 0.296 & 7.99 & -0.80 & 0.76 & 1.18 $\pm$ 0.28 \\
2308250227 & 0.348 & 8.18 & -0.64 & 0.78 & 1.45 $\pm$ 0.34 \\
1502275236 & 0.316 & 8.64 & -0.21 & 0.86 & 0.79 $\pm$ 0.03 \\
2310330106 & 0.283 & 7.98 & -0.53 & 1.05 & 0.94 $\pm$ 0.22 \\
2310184187 & 0.350 & 9.43 & 0.61 & 1.09 & 0.87 $\pm$ 0.21 \\
2302280234 & 0.291 & 8.07 & -0.37 & 1.14 & 0.62 $\pm$ 0.10 \\
2310313103 & 0.284 & 9.67 & 0.86 & 1.16 & 0.91 $\pm$ 0.04 \\
2310037140 & 0.341 & 9.75 & 0.99 & 1.21 & 1.01 $\pm$ 0.16 \\
1533258059 & 0.407 & 8.74 & 0.36 & 1.35 & 0.67 $\pm$ 0.14 \\
\bottomrule
\end{tabular}
}
\end{table*}

\clearpage

\begin{figure*}[ht]
    \centering
    \includegraphics[width=\linewidth]{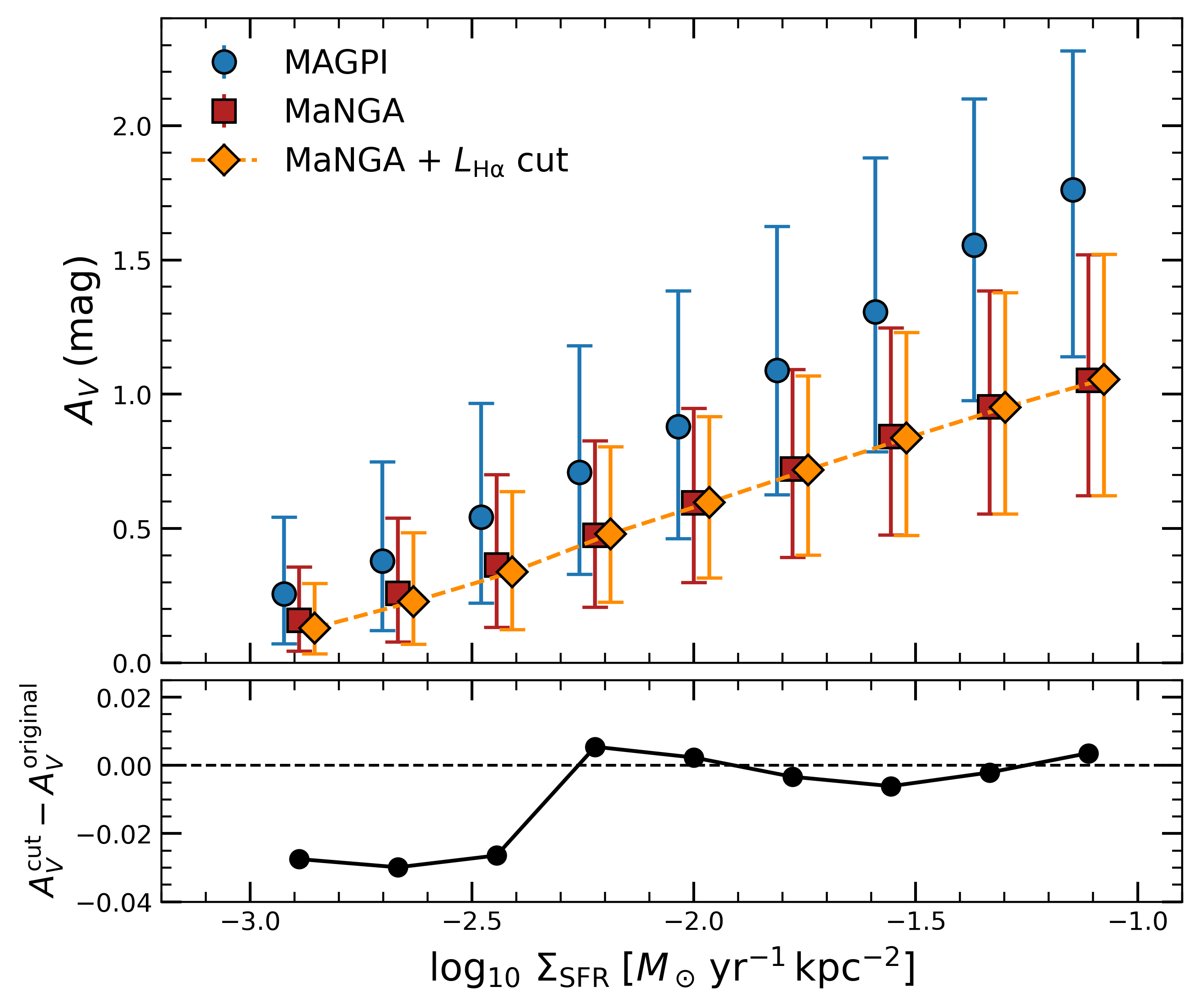}
    \caption{Effect of imposing a MAGPI-like H$\alpha$ luminosity threshold ($L_{\mathrm{H}\alpha,\mathrm{lim}} = 5.6 \times 10^{37}\,\mathrm{erg\,s^{-1}}$) on the MaNGA comparison sample. The top panel shows the median $A_V$--$\Sigma_{\rm SFR}$ relation for the original MaNGA sample and for the MaNGA sample after applying the detection cut, with the MAGPI relation shown for reference. The bottom panel shows the difference between the two MaNGA relations.}
    \label{fig:detection_limit_test}
\end{figure*}

\begin{figure*}[ht]
    \centering
    \includegraphics[width=\linewidth]{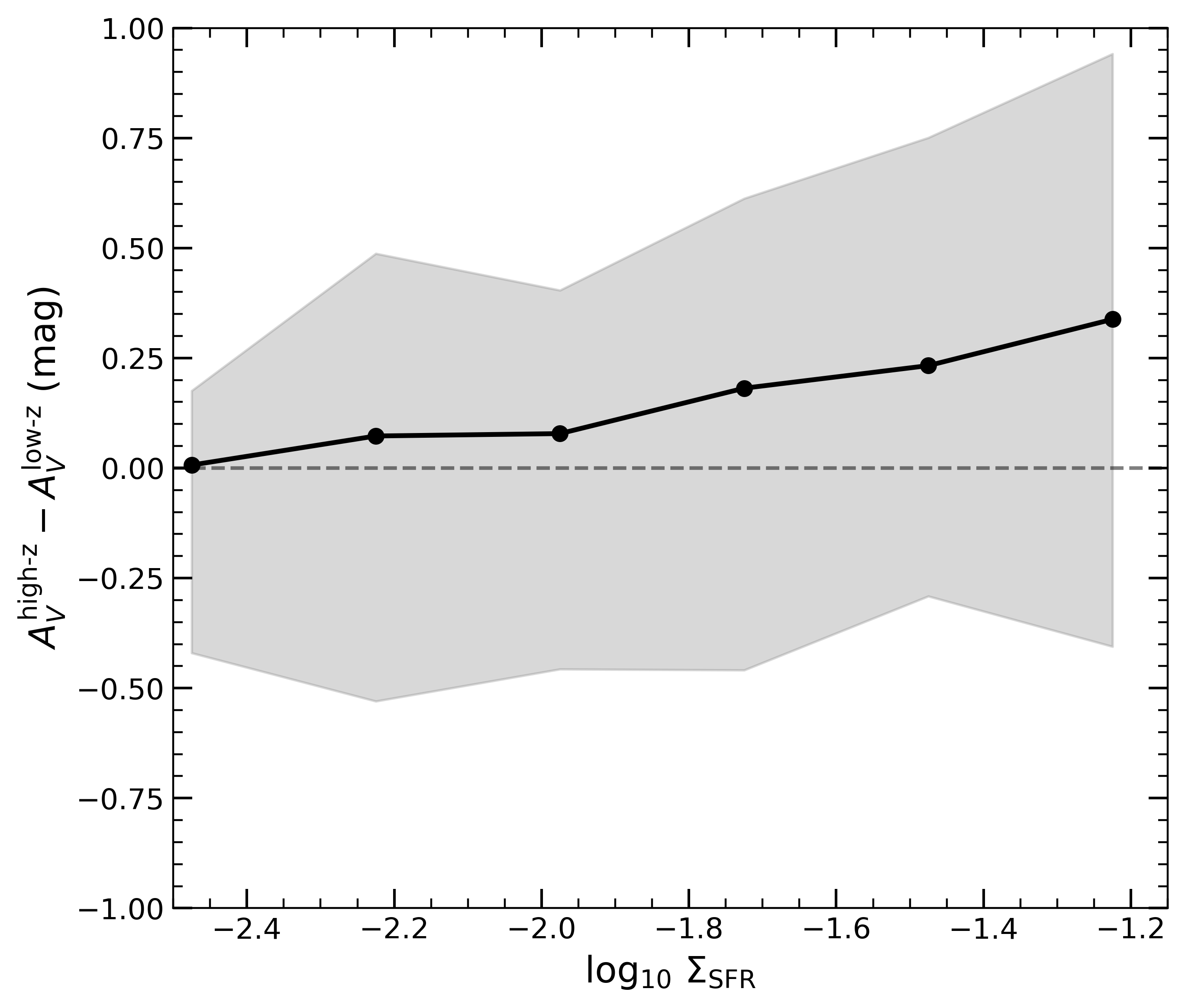}
    \caption{Resolved dust attenuation difference as a function of star formation rate surface density, $\log_{10}\,\Sigma_{\mathrm{SFR}}$, for matched MAGPI galaxies. The attenuation offset is defined as $\Delta A_V = A_V^{\mathrm{high\text{-}z}} - A_V^{\mathrm{low\text{-}z}}$, comparing galaxies in the redshift ranges $0.30 < z < 0.45$ (high-z) and $0.25 < z \leq 0.30$ (low-z). Galaxies are matched in stellar mass and offset from the star-forming main sequence ($\Delta$SFMS). The solid line shows the median $\Delta A_V$ across Monte Carlo realisations, with the shaded region indicating the 16th–84th percentile range. The dashed horizontal line marks $\Delta A_V = 0$.}

    \label{fig:av_evolution}
\end{figure*}

\end{document}